%% file: belayer.tex
\documentclass[sigplan,noacm]{acmart}

\input{packages}
\usepackage[framemethod=tikz]{mdframed}

\begin{document}

\title{Belayer: Efficient Fault Tolerance for LLM Agentic RL Training}

\author{\texorpdfstring{Jiecheng Zhou\textsuperscript{1,2}, Qinghao Hu\textsuperscript{3}, Peng Sun\textsuperscript{4}, Xingcheng Zhang\textsuperscript{2}, Weiming Zhang\textsuperscript{1}}{Jiecheng Zhou, Qinghao Hu, Peng Sun, Xingcheng Zhang, and Weiming Zhang}}
\affiliation{%
  \institution{\textsuperscript{1}University of Science and Technology of China \quad
  \textsuperscript{2}Shanghai AI Laboratory \quad
  \textsuperscript{3}MIT \quad
  \textsuperscript{4}Unaffiliated}
  \city{}
  \country{}}
% Here to avoid show superscript in ACM Reference Format.
\renewcommand{\authors}{Qinghao Hu, Meng Zhang, Peng Sun, Yonggang Wen, and Tianwei Zhang}
\renewcommand{\shortauthors}{}
\newcommand{\SN}{\texttt{Belayer}\xspace}

% \settopmatter{authorsperrow=3}

% \renewcommand{\shortauthors}{Qinghao Hu, Peng Sun, Shengen Yan, Yonggang Wen, and Tianwei Zhang}

\input{0_Abstract}

%   <ccs2012>
%   <concept>
%   <concept_id>10010520.10010521.10010537.10003100</concept_id>
%   <concept_desc>Computer systems organization~Cloud computing</concept_desc>
%   </concept>
%   <concept>
%   <concept_id>10010147.10010178.10010199</concept_id>
%   <concept_desc>Computing methodologies~Planning and scheduling</concept_desc>
%   </concept>
%   </ccs2012>
% Remove ACM running headers, keep page numbers.

% \ccsdesc[500]{Computer systems organization~Cloud computing}
% \ccsdesc[500]{Computing methodologies~Planning and scheduling}

\keywords{LLM Reinforcement Learning, Fault-tolerance}

\maketitle

\input{1_Introduction}
\input{2_Motivation}
\input{3_System_Design}
\input{4_Rollout_FT}
\input{5_Env_FT}
\input{6_Cluster_FT}
\input{7_Evaluation}

\input{8_discussion}
\input{9_Conclusion}

% \newpage
\bibliographystyle{ACM-Reference-Format}
\bibliography{references}

\input{Appendix}

% 如果不需要强制单栏多栏控制，可以删除以下重复部分
% \onecolumn
% \begin{multicols}{2}
%   \bibliographystyle{ACM-Reference-Format}
%   \bibliography{references}
% \end{multicols}

\end{document}

%% file: packages.tex
\usepackage{xspace}
\usepackage{mathtools} % upgrade from amsmath
\usepackage{bm}
\usepackage{enumitem}
\usepackage{textcomp}
\usepackage{xcolor}
\usepackage{changepage}

\usepackage{array}
\usepackage{ragged2e}

\usepackage{graphicx}
\usepackage{subcaption}
\usepackage{tikz}

\usepackage{tabularx}
\usepackage{multirow}
\usepackage{booktabs}
\usepackage{makecell}
\usepackage[figuresleft]{rotating}
\usepackage[flushleft]{threeparttable}

\usepackage{algorithmicx}
\usepackage{algorithm}
\usepackage[noend]{algpseudocode}

\algnewcommand\algorithmicinput{\textbf{Input:}}
\algnewcommand\Input{\item[\algorithmicinput]}
\algnewcommand\algorithmicoutput{\textbf{Output:}}
\algnewcommand\Output{\item[\algorithmicoutput]}

\usepackage[utf8]{inputenc} % Avoid insert Chinese characters
\usepackage{listings}
\usepackage{tcolorbox} % Colored and framed text boxes
\usepackage{soul} % Colored and framed text 
\usepackage{pifont} % inconsolata
\usepackage{bbding} % icons
\hypersetup{hidelinks}
\newcommand{\onex}{\raisebox{-0.6mm}{\large{\ding{172}}}}
\newcommand{\twox}{\raisebox{-0.6mm}{\large{\ding{173}}}}
\newcommand{\threex}{\raisebox{-0.6mm}{\large{\ding{174}}}}
\newcommand{\fourx}{\raisebox{-0.6mm}{\large{\ding{175}}}}
\newcommand{\fivex}{\raisebox{-0.6mm}{\large{\ding{176}}}}
\newcommand{\sixx}{\raisebox{-0.6mm}{\large{\ding{177}}}}
\newcommand{\sevenx}{\raisebox{-0.6mm}{\large{\ding{178}}}}
\newcommand{\eightx}{\raisebox{-0.6mm}{\large{\ding{179}}}}

%% file: 0_Abstract.tex
\begin{abstract}

Large language model (LLM) agents are increasingly trained with reinforcement learning in long-horizon, sandboxed environments. 
Unlike conventional RL, agentic RL couples GPU-intensive rollout engines with stateful environment containers whose actions may produce visible side effects, such as file edits, command execution, and dependency installation. 
A single trajectory can span many rounds of generation and environment interaction, so a component failure can discard completed work or expose the model to an environment state that is inconsistent with its context. 
However, existing systems lack efficient and correct recovery mechanisms for this distributed execution model.
This paper presents \SN, an efficient fault-tolerant system for LLM agentic RL training. 
\SN handles failures in both rollout engines and environment execution while targeting low failure-free overhead. 
For scoped worker-local rollout failures, \SN equips each pre-initialized shadow worker with a selective GPU-state reuse protocol that retains independently owned weights and raw KV-arena allocations after owner and GPU health checks, reinitializes worker-local state, and rebuilds request-specific KV contents from logged token prefixes. 
For environment failures, \SN introduces \texttt{full\_checkpoint} and \texttt{full\_restore} to jointly capture and restore container file-system and runtime state, and coordinates the recovered environment with the LLM context to preserve prefix consistency. 
An adaptive policy opportunistically overlaps full-state checkpointing with natural LLM inference bubbles when the predicted interval is long enough. 
Empirical results show low measured overhead during failure-free training, a worker-recovery-time reduction of up to 42$\times$ compared with a full engine cold start, and 1.5$\times$--3.5$\times$ faster recovery from environment failures.

\end{abstract}

%% file: 1_Introduction.tex
\section{Introduction}
\label{intro}

Large language models (LLMs)~\cite{DeepSeekv4,glm5,openai_gpt5_2025,kimi2.5} are increasingly trained not only to generate text but also to act as agents that solve complex tasks in real world.
They use tools, execute commands, modify files, and iteratively solve long-horizon tasks~\cite{swe-bench,terminal_bench,swe-gym,openhands}.
Reinforcement learning (RL)~\cite{RL} is a natural training paradigm for such workloads because models can learn from trajectories collected through realistic interactions.
In a typical LLM agentic RL pipeline~\cite{OpenClaw_RL}, training alternates between rollout generation, reward computation, and policy optimization~\cite{GRPO,deepseek_r1}. 
During rollout, the agent repeatedly invokes LLM inference, executes an action, observes environment feedback, and updates its context until the trajectory terminates~\cite{ReAct}.
The resulting trajectories are scored and used to update weights.

\begin{figure}[t]
    \centering
    \includegraphics[width=0.85\columnwidth]{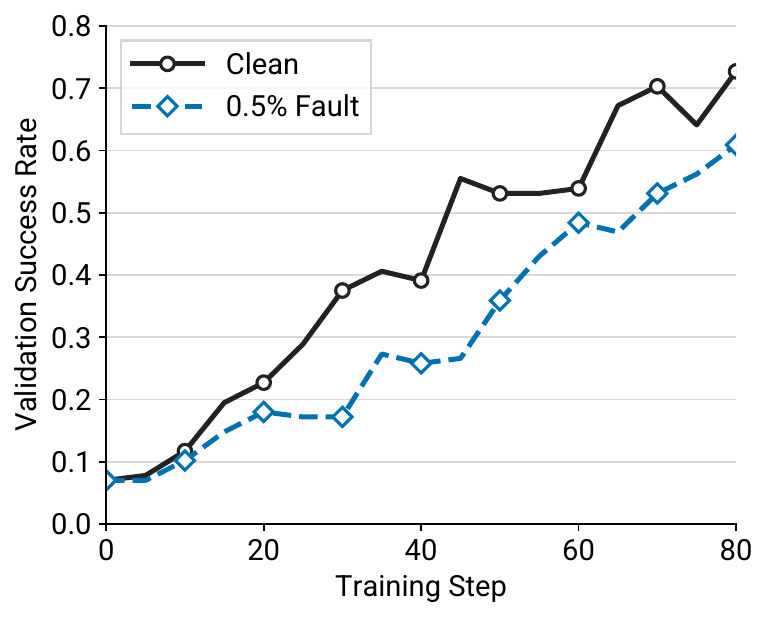}
    \caption{Fault injection results.}
    % (a) Conventional recovery after a rollout-worker failure requires failure detection, reconstruction of the rollout engine, and regeneration of in-flight requests. (b) An infrastructure-induced OOM is surfaced as task feedback and misleads the LLM into taking an irrelevant follow-up action.}
    \label{motivation}
\end{figure}

This emerging workload imposes new system requirements on LLM RL training.
Unlike text-only RL and many simulator-based RL workloads~\cite{RLlib,gymnasium}, 
LLM agentic RL executes actions with persistent \textbf{side effects} inside isolated environments (e.g., Docker containers~\cite{Docker}, virtual machines~\cite{VM}). 
A single trajectory may span many rounds of token generation and environment execution.
Its semantic state spans the logged LLM context and sandbox-local file-system and runtime state.
Rollout engines and environment containers host this state.
At the same time, training jobs run for long periods at cluster scale~\cite{Grok4,kimi2.5}, 
exposing them to software failures and hardware faults documented in training and container deployments~\cite{Acme,RobustLLMTraining_bytedance,yu2024bugs,schroeder2009large,RaidServe,ghostserve}.
To study the impact of failures, we inject faults into a real-world LLM agentic RL workload and measure their effect on training efficiency (Fig.~\ref{motivation}).
We use the setting and source code of GiGPO~\cite{GiGPO}, injecting faults at 0.5\% probability.
Although the injection rate is low, the long interaction steps of trajectories amplify the effect of each failure (about 8.98\% trajectories), as a result, the validation curves are visibly affected.
% As a result, fault tolerance is no longer a peripheral concern; it directly affects training efficiency and trajectory correctness.

Existing fault-tolerance mechanisms are inefficient for this setting. 
Fault tolerance in LLM training has traditionally focused on pretraining~\cite{ma2026resihp,Elaswave,Oobleck,torch-FT,Recycle,Parcae,RobustLLMTraining_bytedance}, where systems periodically checkpoint model state~\cite{ByteCheckpoint,CheckFreq,Gemini,TorchTitan,DataStatesLLM} and restart from the latest checkpoint after a failure.
Although effective for pretraining with short iterations, this approach is too coarse-grained for LLM agentic RL.
In common agentic RL workloads, a training step can take tens of minutes. 
% A job-wide restart after one component fails reinitializes whole training processes including the rollout engines and training workers, and regenerates all trajectories, wasting computation and slowing training.
More recent systems~\cite{Role-based_FT} adopt role-based recovery to isolate failures to the affected component. 
However, even rollout engine recovery remains costly\cite{laminar} and, more importantly, does not ensure that the recovered LLM context and environment state correspond to the same trajectory prefix.
A conventional recovery after environment failures is restarting the whole trajectory, which can be inefficient. 
This is because single trajectory typically takes a long time to complete.
In our measurement, the median trajectory lasts 460.8 seconds (Fig.~\ref{fig:rollout_analysis} (b)), and 1\% of environment action failures can extend the rollout duration by 48.7\%.
% Restarting a rollout engine discards in-flight requests and requires re-initializing heavyweight inference state. 
% Retrying a failed environment action can silently corrupt the environment because tool actions can be non-idempotent\cite{critique_remote_procedure_call}. 
% Dropping failed trajectories wastes computation and may bias the collected experience\cite{DeepSeekv4}.

Together, these limitations leave two gaps for LLM agentic RL.
First, isolating a rollout engine failure does not eliminate the heavyweight reconstruction needed for recovery.
Second, environment recovery either discards completed interactions and restarts the trajectory or lacks an aligned recovery point spanning the LLM context and environment state.

Faced with these gaps, we present \SN, an efficient fault-tolerant system for LLM agentic RL training.
\SN uses a pre-initialized standby worker for rollout recovery~\cite{Primary_backup} and checkpoint/restore full states for environment recovery~\cite{criu,docker_docs}.
A standby moves most heavyweight inference-engine initialization off the failure path (handover within 1 second), while a checkpoint preserves completed interactions and provides a known state from which a trajectory can regenerate after an ambiguous action failure~\cite{critique_remote_procedure_call,implementing_RPC,lee2015implementing}.
% Applying these mechanisms efficiently requires determining which GPU state the standby can retain and when a full environment checkpoint can run without delaying the rollout.

\textbf{Selective GPU-state reuse for standby recovery.}
A pre-initialized standby moves worker initialization off the failure path\cite{REINFORCE}, but keeping it ready to serve requires access to the rollout engine's model-scale GPU state.
Replicating the weights, KV-cache arena, and CUDA-graph buffers for each standby consumes scarce GPU memory and reduces serving capacity.
Sharing the active worker's state wholesale avoids this duplication, but couples recovery to allocations owned by the failed CUDA context and may carry partially updated request-specific KV contents across the failure boundary.
Practical standby recovery therefore requires a finer-grained state boundary that preserves reusable GPU backing state while excluding volatile worker and request state.
% The challenge is to distinguish reusable backing allocations from invalid worker and request state.
Our analysis of real-world inference engine failures (Sec.\ref{opportunities}) yields a key observation: \emph{model weights and the raw KV-cache arena are frequently reuse candidates, whereas process context and complete request-specific KV contents generally are not.}
Based on this observation, \SN places weights and the raw KV arena under independent owners and maps them into a pre-initialized shadow worker.
After a worker failure, the shadow reuses the retained allocations, remaps its graph buffers and rebuilds KV cache from prefix tokens.

\textbf{LLM-response-aware environment checkpointing.}
Environment checkpointing is not a free operation, as it involves freezing the environment, capturing its state, and persisting it.
Our key observation is that agentic rollouts alternate between environment execution and LLM generation: after an action completes, no new agent action is dispatched while the LLM thinks about its next response\cite{ReAct}.
This response-generation interval provides a natural window in which to freeze and checkpoint the environment state without delaying agent execution.
Exploiting the window is non-trivial because the response duration varies, so a checkpoint that outlasts the window delays the next environment action.
Based on this observation, \SN jointly captures file-system and runtime state at a completed action boundary while the LLM generates its next response.
An online risk- and bubble-aware policy launches the checkpoint only when its expected regeneration benefit justifies the latency likely to remain visible.
Each completed checkpoint is atomically associated with the corresponding LLM context, so recovery restores all states from the same ready action boundary.

% \begin{figure}[t]
%     \centering
%     \includegraphics[width=1\columnwidth]{motivation_restart_overhead.pdf}
%     \caption{Naive recovery methods in agentic RL. 
%     All tests use Qwen3-32B on 4 H200 GPUs with TP=4. We use 0.01 failure rate at each environment action in figure (a).}
%     % (a) Conventional recovery after a rollout-worker failure requires failure detection, reconstruction of the rollout engine, and regeneration of in-flight requests. (b) An infrastructure-induced OOM is surfaced as task feedback and misleads the LLM into taking an irrelevant follow-up action.}
%     \label{restart_overhead}
% \end{figure}

% As a practical implementation optimization, \SN also uses historical resource profiles and current host headroom to limit concurrent environment container launches.
% TODO: modification
We implement \SN on Slime with $\sim$6K lines of code and evaluate Qwen3 models ranging from 4B to 32B parameters on math-reasoning and software-engineering (SWE) workloads.
\SN has low failure-free overhead, achieves up to a 42$\times$ worker-recovery speedup over a full engine cold start, and provides a 1.5$\times$--3.5$\times$ recovery speedup over the environment failure baseline.
% The code will be released upon publication.

In summary, we make the following contributions:
\begin{itemize}[itemsep=0pt, topsep=1pt, leftmargin=*]
\item We design a selective GPU-state reuse protocol for shadow workers that retains independently owned backing allocations after health checks while reconstructing volatile worker and request state.
\item We design an LLM-response-aware environment checkpointing mechanism that overlaps full-state capture with response generation and uses an online risk-aware policy to checkpoint only when its expected recovery benefit justifies the exposed latency.
\item We implement \SN on Slime in $\sim$6K lines and evaluate Qwen3 models from 4B to 32B parameters. Evaluation shows low failure-free overhead and considerable performance improvements.
\end{itemize}

%% file: 2_Motivation.tex
%-------------------------------------------------------------------------------
\section{Background and Motivation}
\label{sec_background}
%-------------------------------------------------------------------------------
\subsection{Agentic RL Workloads and Recovery State}
\label{agentic_rl_background}

Similar to text-only RL~\cite{ouyang2022training}, agentic RL trains LLMs in common RL workflow:
1) Rollout: It firstly generates a batch of trajectories, each consisting of a sequence of LLM-generated tokens and environment observations.
2) Rewarding: It then computes a reward for each trajectory.
3) Model training: It updates the model parameters based on the trajectories and their rewards, finally producing a new policy version for the next rollout.

During an agentic RL rollout, the model alternates between token generation and tool execution in a sandbox.
Unlike text-only RL~\cite{ouyang2022training}, tool actions typically modify persistent file-system and runtime state, and their observations shape later actions and rewards.
Consequently, a valid trajectory depends on both its logged LLM context and the corresponding environment state.

At a committed action boundary $k$, let $L_k$ be the logged LLM context and $E_k=(F_k,R_k)$ the sandbox-local environment state, where $F_k$ and $R_k$ are the file-system and runtime components.
A consistent trajectory prefix is
\begin{equation}
    P_k=(L_k,E_k)=(L_k,F_k,R_k).
\end{equation}
A prefix becomes ready for environment recovery \textbf{only after $F_k$ and $R_k$ have been captured and associated with $L_k$.}
% During an in-flight response, $E_k$ remains unchanged while the rollout router records a finer-grained token prefix for rollout-worker recovery.
% Worker processes and GPU backing allocations belong to the recovery substrate rather than $P_k$; the policy version and inference configuration are fixed rollout metadata.

\emph{Why no external state?}
% External states, such as remote services or host-mounted volumes, are excluded.
Because model-generated actions are untrusted, \SN targets deployments whose sandboxes prevent writes outside the container~\cite{Docker,VM,swe-gym,OpenClaw_RL}.
Mutable task state is therefore confined to $(F_k,R_k)$; deployments permitting writable remote services or host-mounted volumes fall outside this guarantee.
% Within this scope, recovery assumes that the initial image, trajectory log, fixed policy version, and latest ready checkpoint remain accessible.

\subsection{Failure Consequences and Recovery Gaps}

Production studies document node, accelerator, communication, process, and container failures in large-scale training and cluster execution~\cite{Acme,RobustLLMTraining_bytedance,schroeder2009large,verma2015borg,yu2024bugs,simonsson2019chaosorca}.
Agentic RL runs atop the same fallible infrastructure, but its stateful and interactive execution model turns familiar failures into new recovery problems.

First, a rollout-worker failure is more than a transient loss of serving capacity: it interrupts stateful in-flight requests.
In our Qwen3-32B setup, conventional recovery spends approximately 30 seconds detecting the failure through the heartbeat, 38.5 seconds reconstructing the rollout engine, and approximately tens of seconds regenerating interrupted requests.
Thus, even a worker failure can add considerable delays to the affected trajectory's generation.

Second, a failure in the environment infrastructure can \emph{change the semantics of the training sample} rather than merely delay it.
% An infrastructure error can also be returned to the model as if it were ordinary task feedback.
In the Prime~\cite{Prime} example in Fig.~\ref{motivation}, an injected OOM traceback appears as an ordinary observation.
The LLM mistakes it for a task-level memory problem and changes its next action, leaving the actual semantic bug unexposed.
If a failure interrupts an action, the executor may not know whether the action had no effect, completed successfully, or left partial effects; retrying a non-idempotent action from this uncertain state can further corrupt the trajectory.
% Conventional recovery that restarts the failed trajectory is costly (Fig.~\ref{restart_overhead} (a)).
% This experiment illustrates the consequence, not the production incidence, of exposing infrastructure failures to the agent.
Together, the examples show that familiar infrastructure failures create a new correctness problem in agentic RL.

These consequences imply two independent recovery requirements.
First, recovery must limit critical-path delay: even if healthy trajectories continue, slow recovery can extend the completion tail.
% Restoring service availability alone is insufficient: a failure may discard request progress or leave the LLM context and environment state at different trajectory prefixes.
Second, checkpoint/restore mechanisms must preserve trajectory validity by rolling the LLM context and environment state back to a consistent prefix.
% We evaluate the potential efficiency benefit under measured workload durations and a range of failure rates in Sec.~\ref{trajectory_state_recovery}.
We focus on detected component failures that interrupt rollout or environment execution; silent data corruption, control-plane failures, external mutable effects, and loss of persistent recovery inputs are out of scope.

\begin{table}[t]
\centering
\small
\setlength{\tabcolsep}{3.2pt}
\caption{State coverage within \SN's container-local recovery boundary. Aligned indicates a common legal action-boundary prefix.}
\label{tab:recovery_boundary_comparison}
\begin{tabular*}{\columnwidth}{@{\extracolsep{\fill}}lcccc@{}}
\toprule
\textbf{Primitive/protocol} & $L_k$ & $F_k$ & $R_k$ & \textbf{Aligned} \\
\midrule
Trajectory log & Yes & No & No & No \\
Git repository state & No & Partial & No & No \\
Overlay snapshot & No & Yes & No & No \\
CRIU process checkpoint & No & No & Yes & No \\
\SN ready prefix & Yes & Yes & Yes & Yes \\
\bottomrule
\end{tabular*}
\vspace{0.2em}
\begin{minipage}{0.98\columnwidth}
\footnotesize
% Git covers versioned repository content, but not untracked artifacts, installed dependencies, or live processes.
% The overlay and CRIU rows refer to the individual storage and runtime primitives, respectively; $R_k$ denotes only runtime state that CRIU can restore within the stated scope.
\end{minipage}
\end{table}

\textbf{Existing work.}
Prior training fault tolerance typically checkpoints model and optimizer state and restarts the job after a failure~\cite{ByteCheckpoint,CheckFreq,Gemini,TorchTitan,DataStatesLLM,RobustLLMTraining_bytedance}.
Role-based recovery narrows the restart to the affected training role~\cite{Role-based_FT}, but still incurs significant overhead.
% Serverless LLM-serving systems reduce this startup cost through faster weight loading, runtime-state materialization, and proactive model distribution~\cite{ServerlessLLM,MedusaMaterialization,HydraServe}.
% However, they target engine provisioning rather than failover rollout requests or aligning them with environment state.
Similarly, retaining generated tokens in a trajectory log can reconstruct $L_k$, but not the environment state that produced those tokens.
Table~\ref{tab:recovery_boundary_comparison} summarizes the resulting state-boundary gap.
Version control suffices when the required state consists only of explicitly committed source code, but agent actions can also create untracked files, build artifacts, installed packages, and long-lived processes.
An overlay snapshot captures the container's writable file-system layer~\cite{docker_storage_drivers}, whereas CRIU captures restorable process state~\cite{criu}; either component alone can be inconsistent with the other.
A recovery protocol must still capture $F_k$ and $R_k$ at one quiescent point, publish them atomically as a ready checkpoint, and associate that checkpoint with $L_k$.

\subsubsection{Standby Recovery for Rollout Engines}
\label{rollout_ft_background}
A cold-start replacement cannot serve requests until it has rebuilt the inference stack and initialized its GPU state, placing this entire delay on the critical path.
A pre-initialized standby performs worker-local initialization before a failure and remains ready to take over.
Primary-backup coordination can then redirect requests to the standby after the active worker fails~\cite{Primary_backup,Chubby}.
This removes most worker initialization from recovery's critical path.

\noindent\textbf{Why direct application is insufficient.}
The GPU state that makes inference-engine initialization expensive also makes a conventional standby impractical.
Giving every shadow worker independent weights, KV-cache capacity, and CUDA-graph buffers replicates model-scale GPU state.
Sharing all of the active worker's state avoids this duplication, but crosses the failure boundary: some failures are tied to the failed CUDA context, and request-specific KV contents may be only partially updated when the worker fails.
Thus, neither full replication nor indiscriminate sharing provides an efficient and safe recovery path.

\noindent\textbf{Challenge.}
Existing inference engines treat each worker as an indivisible recovery unit: model weights, KV cache, runtime objects, and request state reside in one process/CUDA context and share its lifetime.
The engine therefore exposes only all-or-nothing recovery---keeping the context retains every state, including potentially invalid or partially updated state, whereas destroying it discards every state and requires full reconstruction.
The challenge is to break this atomic recovery boundary without allowing potentially invalid state to cross the failure boundary.

\begin{table}[t]
  \caption{Reusable candidates in the fail-stop issue corpus.}
%   An issue is a reusable candidate only when no identified failure path in that
%   report implicates the state.}
  \label{tab:reuse-opportunity}
  \centering
  \normalsize
  \setlength{\tabcolsep}{3.5pt}
  \renewcommand{\arraystretch}{1.05}
  \begin{tabular}{@{}lrrrrr@{}}
    \toprule
    Reuse candidate & vLLM & SGLang & TRT-LLM & TGI & Total \\
    \midrule
    Process context \(P\)                 & 1  & 0  & 0  & 0 & 1  \\
    Weights \(W_{\mathrm{mem}}\)          & 16 & 15 & 13 & 5 & 49 \\
    KV arena \(A_{\mathrm{mem}}\)         & 16 & 16 & 14 & 6 & 52 \\
    Complete KV \(K\)                     & 0  & 7  & 2  & 0 & 9  \\
    \(W_{\mathrm{mem}}\cap A_{\mathrm{mem}}\) & 16 & 14 & 13 & 5 & 48 \\
    \bottomrule
  \end{tabular}
\end{table}

\subsubsection{Environment State Recovery}
% As shown in Table~\ref{tab:recovery_boundary_comparison}, no individual checkpointing primitive provides an aligned recovery prefix.
% A trajectory log captures only the LLM context, while a writable-layer snapshot and a runtime checkpoint each cover one environment component.
% Moreover, after a failure interrupts an action, replaying that action from an uncertain environment state is unsafe because tool actions can be \emph{non-idempotent}.
Checkpoint/restart mechanisms for agentic RL must capture the container's file system and runtime state at the same committed action boundary, associate this state with the corresponding LLM context and continue the rollout process.

\textbf{Challenge.}
Satisfying both correctness and efficiency requirements is non-trivial.
Capturing state while an action is in flight can mix file-system and runtime effects from different trajectory prefixes, while a failure during checkpoint creation can leave an incomplete recovery point.
Moreover, jointly checkpointing runtime and file-system state is expensive: frequent checkpoints increase failure-free latency, whereas sparse checkpoints increase rollback and regeneration costs after a failure.
The key challenge is therefore to preserve prefix consistency while keeping visible failure-free overhead negligible.

\subsection{Opportunities}
\label{opportunities}
\begin{figure}[t]
    \centering
    \begin{subfigure}[b]{1\columnwidth}
        \centering
        \includegraphics[width=1\columnwidth]{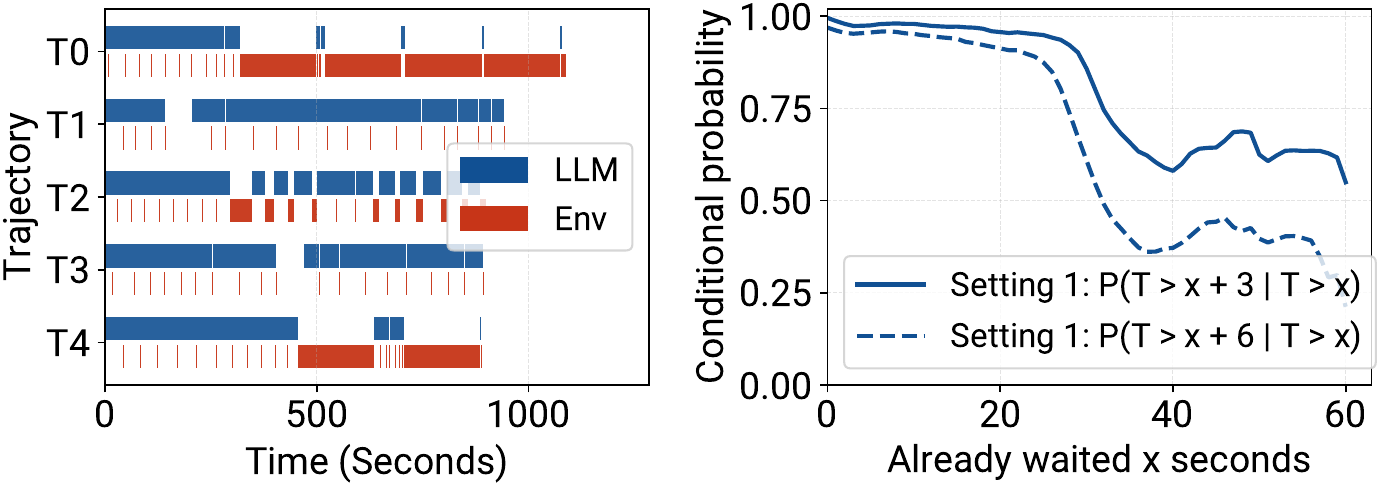}
        \caption{Sample timelines and the conditional probability that an ongoing LLM response has at least 3 or 6 seconds remaining.}
        \label{trajectory_timing}
    \end{subfigure}

    \begin{subfigure}[b]{0.9\columnwidth}
        \centering
        \includegraphics[width=1\columnwidth]{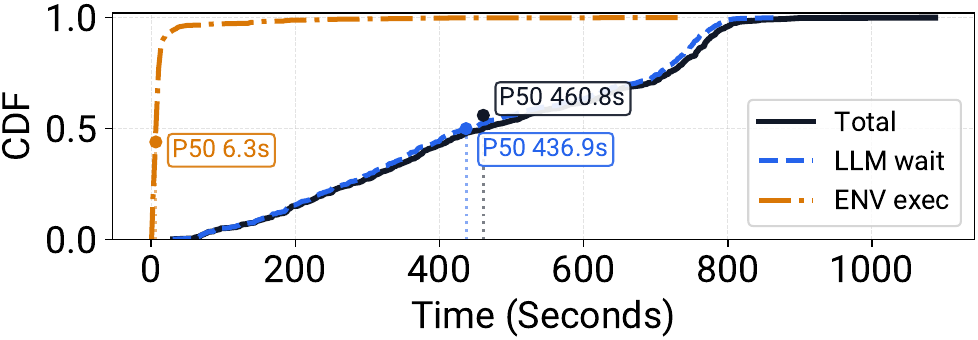}
        \caption{CDF of per-trajectory total time and cumulative time spent waiting for LLM responses and executing environment actions.}
        \label{breakdown}
    \end{subfigure}
    \caption{Measured rollout timing and workload-dependent opportunities to overlap environment checkpointing with LLM generation.}
    \label{fig:rollout_analysis}
\end{figure}

\noindent\textbf{Reuse across failure boundaries.}
To determine what a standby can safely retain, we ask which logical states are
not implicated by reported failures and could therefore be reused if placed
under an independent owner outside the worker's lifetime.
We search the public GitHub issue trackers of vLLM, SGLang, TensorRT-LLM, and
Text Generation Inference (TGI) over a fixed three-year window, covering
27,644 pages, and filter them using severe-impact keywords for crashes, hangs,
and fatal CUDA errors.
This audit yields 5,834 deduplicated candidate issue pages.
After deduplicating aliases, we randomly select 58 issue reports---21 from vLLM, 17
from SGLang, 14 from TensorRT-LLM, and 6 from TGI---for the reuse analysis.
Each report is (i) a public user report, (ii) a failure in an
inference-serving path rather than startup, build, configuration, or model
download, (iii) associated with an observable fail-stop outcome, including a
crash, worker exit, fatal CUDA error, resource-exhaustion termination, or
externally visible hang or wedge, and (iv) supported by enough issue, fix, or
code-path evidence to determine the implicated state.

For each report, we use linked fixes or root-cause analyses to identify the
implicated logical state.
As shown in Table~\ref{tab:reuse-opportunity}, we distinguish the process
context \(P\), which contains CPU process state and its CUDA context, from
weight memory \(W_{\mathrm{mem}}\), the raw KV-cache arena
\(A_{\mathrm{mem}}\), and the complete request-specific KV image \(K\).
We classify a state as reusable only if no identified failure path in the
report implicates it.
A reusable state means only that available evidence does not show corruption of
that state's logical contents.
% It does not imply that an allocation survives process exit or is safe to reuse without independent ownership and post-failure validation.

\begin{figure*}[t]
    \centering
    \includegraphics[width=0.95\textwidth]{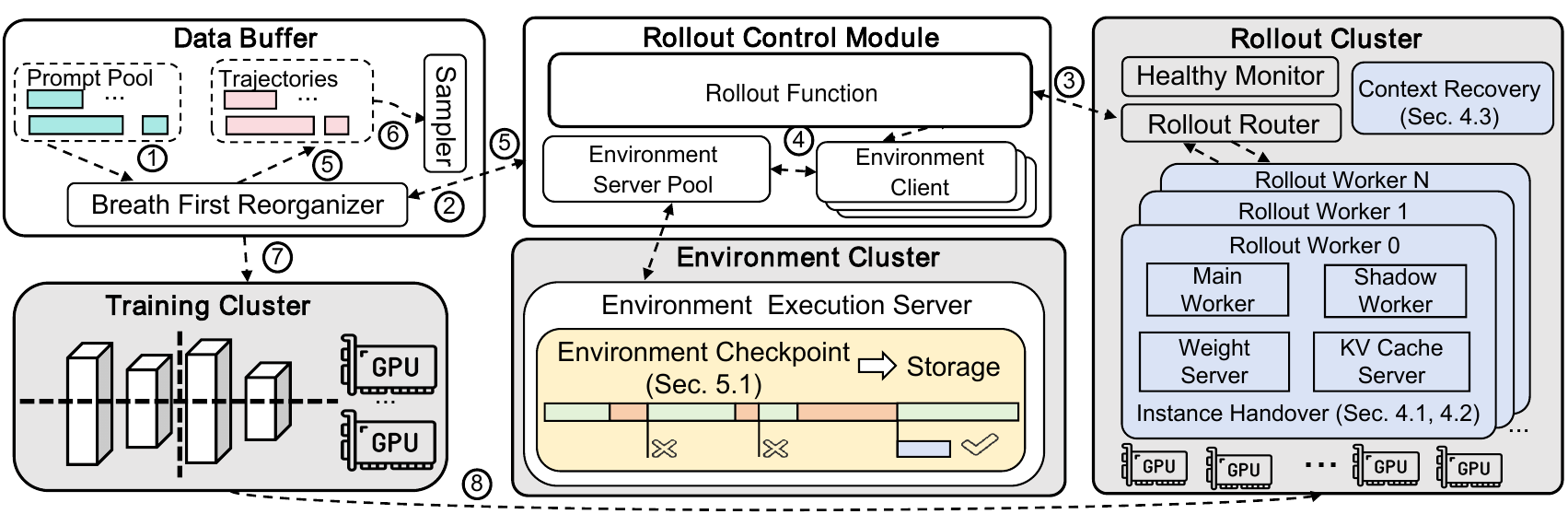}
    % \caption{Training workflow and system overview. Workload-aware admission is an optional implementation optimization; the core recovery paths are rollout handover and environment checkpointing.}
    \caption{Training workflow and system overview.}
    \label{fig_system_overview}
\end{figure*}

\emph{Finding 1: Weight and KV-arena contents are frequently reuse candidates in this corpus.}
For 48 of 58 (82.8\%) fail-stop reports, neither model weights nor the raw KV
backing arena is implicated, making both candidates for reuse under the stated
conditions.

\emph{Finding 2: Process context and complete KV contents are rarely reuse candidates.}
Only one of 58 (1.7\%) reports leaves the process context unimplicated, and
only 9 of 58 (15.5\%) leave the complete KV image reusable.
Complete KV caches are often not reusable after failures because ownership and
commit boundaries become ambiguous, making metadata and cached contents
impossible to verify as mutually consistent.
For example, if a worker fails while writing a batch's KV entries, recovery
cannot reliably distinguish fully committed entries from partially written ones.

% OOMs/leaks/assertions/worker deaths, 
\textbf{Generation-time overlap.}
After an action finishes, its environment waits for the next LLM response, creating an opportunity to move checkpoint work off the rollout critical path.
% As shown in Figure~\ref{fig:rollout_analysis}, 
% the median trajectory lasts 460.8 seconds, with 436.9 seconds spent waiting for LLM responses but only 6.3 seconds executing environment actions (Figure~\ref{breakdown}).
% However, response intervals vary, so aggregate waiting time does not guarantee that every checkpoint can be hidden.
Figure~\ref{trajectory_timing} reports the conditional probability that an ongoing response has enough time remaining to hide a checkpoint.

%% file: 3_System_Design.tex
\section{System Overview}
\label{sec_system_overview}

We present \SN, an efficient fault-tolerance system for LLM agentic RL training that combines fast rollout recovery with prefix-consistent environment-state recovery.
% The implementation also provides lightweight admission control as an optional execution optimization.
Figure~\ref{fig_system_overview} shows \SN's control and execution planes.
The control plane manages rollout logic, the training pipeline, and data processing, while the execution plane runs rollout, environment, and training workloads.
Loosely coupled components support flexible deployment and fault isolation.
The control plane has two components:
\begin{itemize}[itemsep=1pt,leftmargin=*]
\item \emph{Data Buffer} loads prompts, post-processes trajectories, and persists the results.
\item \emph{Rollout Control Module} orchestrates rollouts~\cite{agent_survey}.
% It may also apply the optional admission control in Sec.~\ref{sec:admission_optimization}.
\end{itemize}
\noindent The execution plane consists of three components:
\begin{itemize}[itemsep=1pt, leftmargin=*]
\item \emph{Rollout Cluster} receives requests from the rollout router and conducts LLM inference.
\item \emph{Environment Cluster} executes environment interactions in sandboxes.
\item \emph{Training Cluster} performs policy optimization using the trajectories collected by the rollout phase.
\end{itemize}

\noindent \textbf{System interface.} \SN extends Slime's API and training logic~\cite{slime_github} while preserving compatibility.
% users implement a custom rollout function to define the interaction logic of the LLM and the environment. 
% Users can enable \SN's fault-tolerance and execution optimizations with only a few configuration changes.
Users can enable \SN's fault-tolerance mechanisms with only a few configuration changes.

\noindent \textbf{Training workflow.} In Figure~\ref{fig_system_overview}, black arrows show the training workflow.
After initialization, the breadth-first reorganizer loads the dataset, traverses the prompts in each batch (\onex), and passes them to the custom rollout function (\twox).
For each prompt, the rollout function creates an environment container and invokes the LLM inference API (\threex) and environment interaction API (\fourx) to generate a trajectory.
A router dispatches inference requests to rollout engines, while a client--server interface abstracts environment details.
After enough trajectories have been generated, the reorganizer groups them by prompt (\fivex).
The sampler computes rewards and converts the reordered trajectories into training experience (\sixx), which the training cluster uses for policy optimization (\sevenx).
The updated weights then return to the rollout cluster for the next training iteration (\eightx).

\SN provides fault tolerance across the rollout, environment, and training clusters. 
The rollout and environment clusters are the primary focus; their recovery mechanisms are described in Sec.~\ref{rollout-fault-tolerance} and Sec.~\ref{environment-fault-tolerance}, respectively.

For the training cluster, \SN adopts existing LLM pretraining fault-tolerance techniques and isolates learner failures from the rest of the pipeline.
At the end of each training step, \SN asynchronously checkpoints model parameters and optimizer state~\cite{TorchTitan}.
Because agentic rollouts are long-running, this overhead can overlap with rollout generation in the next step~\cite{zhou2025rlwildcharacterizingrlvr}.
After a training failure, \SN restores the learner from the latest checkpoint and resumes at the step boundary.

Overall, \SN provides an end-to-end fault-tolerance framework for LLM agentic RL training, covering failures across rollout, environment, and training components.

%% file: 4_Rollout_FT.tex
\section{Fast Rollout-Worker Recovery}
\label{rollout-fault-tolerance}

This section presents \SN's protocol for fast recovery from detected \emph{fail-stop} failures and stalls in rollout engines.
We first describe selective reuse of persistent GPU-resident backing allocations and the preparation of a shadow worker with fresh volatile state (Sec.~\ref{shadow-worker}).
We next explain the failure-triggered handover protocol (Sec.~\ref{handover}).
Finally, we present fine-grained LLM context recovery, which avoids expensive regeneration of interrupted requests (Sec.~\ref{llm-context-recovery}).

\textbf{Failure model.}
\SN's warm handover targets detected fail-stop failures and stalls confined to the rollout worker, such as Python exceptions, process crashes, and worker hangs.
Silent data corruption (SDC) and fail-slow behavior are outside this scope.
\vspace{-5pt}
\subsection{Selective State Reuse and Shadow Worker}
\label{shadow-worker}

\begin{figure}[t]
    \centering
    \includegraphics[width=0.9\columnwidth]{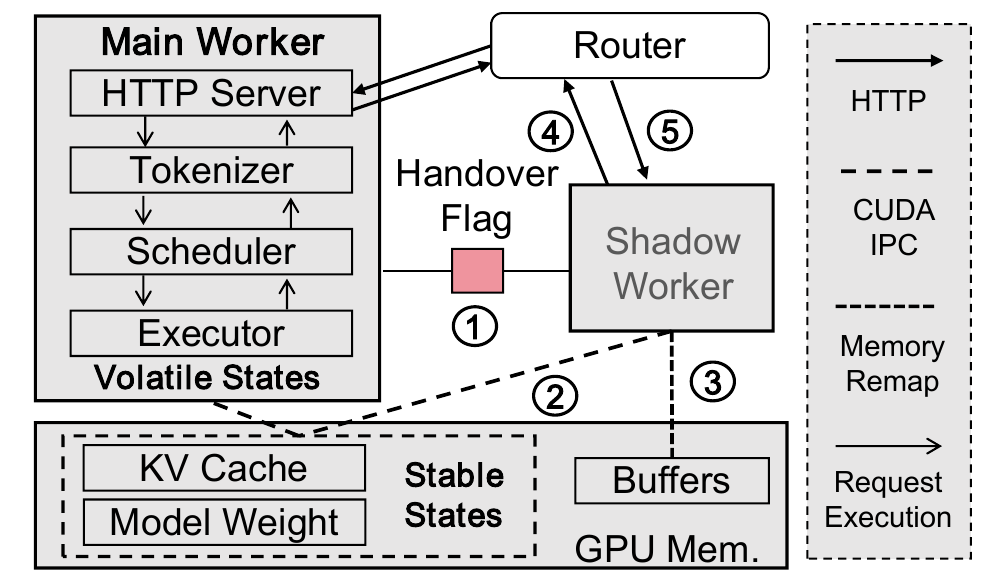}
    \caption{Rollout-engine handover.}
    \label{rollout_restart}
\end{figure}

\textbf{State isolation.}
As Finding~1 in Sec.~\ref{opportunities} indicates, a worker-local failure can invalidate volatile execution state while leaving independently owned GPU-resident allocations as reuse candidates.
\SN therefore moves model weights and the KV-cache memory pool into independent weight and KV-cache servers, which do not execute inference and expose CUDA IPC handles to rollout workers.
These allocations are reused only if their owner processes remain alive and the GPU passes a post-failure health check.
The replacement worker reconstructs worker-local state, including communication groups, the CUDA runtime, request-scheduler state, CUDA-graph buffers, and serving processes.

\textbf{Shadow-worker preparation.}
After isolating these long-lived allocations, \SN pre-initializes a shadow worker for each rollout engine before rollout begins. 
The shadow worker accesses the independently owned backing allocations through CUDA IPC handles and initializes its own volatile execution state.
Colocating it with the main worker, however, may interfere with normal inference.
To avoid this, \SN uses two mechanisms.
First, once initialized, the shadow worker remains idle except for periodic checks of a handover flag.
It performs no inference, minimizing CPU and GPU interference with the main worker.
Second, while on standby, it unmaps its CUDA-graph placeholder buffers using CUDA VMM~\cite{torch_memory_saver}.
These buffers are remapped only when the shadow worker takes over.
Together, these mechanisms minimize failure-free overhead (Sec.~\ref{e2e_evaluation}).

% \textbf{Weight-update consistency.}
% Rollout engines receive new model weights between training iterations.
% \SN treats weight update as a synchronization boundary.
% The weight server publishes a new epoch only after all shards are updated, and both the main worker check the epoch before serving.

\subsection{Handover}
\label{handover}

With state isolation and shadow workers, \SN can perform fast handover after a scoped failure. 
Figure~\ref{rollout_restart} shows the five-step handover. 
\onex The shadow worker detects the main worker's failure through the handover flag.
\twox It checks GPU health with a lightweight CUDA operation, verifies that the weight and KV-cache servers are alive, and flushes the KV-cache pool.
\threex It remaps the CUDA-graph buffers and becomes ready to serve inference requests. 
\fourx It notifies the rollout router to unregister the failed worker and register itself. 
\fivex The rollout router reroutes requests to the new worker, which resumes them through token-level context recovery and reconstructs the KV cache as needed.
Before the next weight update, \SN also re-establishes communication groups between the rollout engine and training cluster.
If the shadow health check fails, \SN marks the GPU unhealthy and does not reuse its resident state.

\textbf{Fencing.} After handover, the control plane terminates any residual main-worker process, thereby preventing a split-brain condition in which the old and replacement workers serve concurrently~\cite{Primary_backup,Chubby}.

\begin{figure}[t]
    \centering
    \includegraphics[width=0.9\columnwidth]{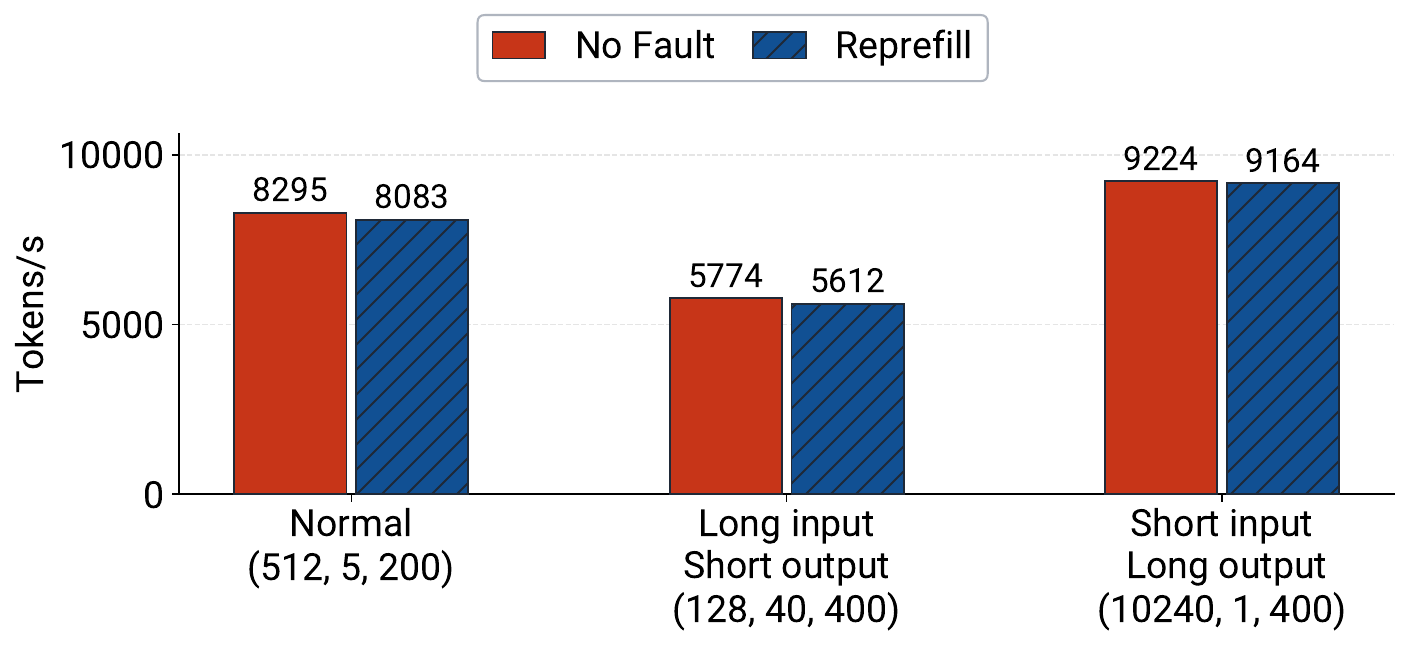}
    \caption{Reprefill cost on GSM8K with Qwen2.5-1.5B and TP=1. Triples in parentheses denote (maximum output tokens, number of few-shot examples, number of requests).}
    \label{reprefill_cost}
\end{figure}

\SN flushes and refills the KV cache during handover for two reasons. 
First, Figure~\ref{reprefill_cost} shows that refill overhead is relatively small in RL workloads.
Second, the complete KV cache pool may be inconsistent after a failure (Finding~2 in Sec.~\ref{opportunities}).
% For example, if the failure occurs during a KV-cache update, the pool may contain a mixture of old and new entries. 
Recovering the KV cache without refilling would require fine-grained consistency tracking, which we leave to future work.

\textbf{Fail-stop detection.}
The shadow worker also enables lightweight failure detection. 
Because the shadow worker is colocated with the main worker, \SN uses the \emph{mtime} of a temporary file as a local heartbeat: the main worker refreshes its timestamp in the scheduling loop, while the shadow worker monitors it.
This avoids relying solely on remote health checks, which can be inefficient for high-performance LLM inference engines.

% Fenced handover. A heartbeat timeout only marks the main worker as suspected and does not authorize the shadow worker to execute. The recovery coordinator first atomically advances the engine to a fencing epoch and instructs the router to
% stop dispatching requests to, and reject all responses from, the previous epoch. It then forcibly terminates all main-worker ranks and waits until their processes are reaped and their CUDA contexts are confirmed to be destroyed. Only
% after the weight and KV-allocation owners remain alive, the GPU health checks succeed, and the retained allocations pass version and integrity validation does Belayer permit warm reuse. The KV arena is logically invalidated and
% reinitialized with a new allocation epoch; request-specific KV contents and worker-local metadata are never reused. For tensor-parallel engines, all shadow ranks reconstruct their communicators and pass an engine-wide readiness barrier
% before the coordinator atomically publishes the new active epoch. If any process cannot be fenced, a CUDA context remains active, or validation fails, Belayer does not activate the colocated shadow worker and instead falls back to a cold
% restart on another healthy GPU. Thus, failure detection affects liveness, whereas fencing and epoch publication preserve the single-writer safety invariant.

\subsection{LLM Context Recovery}
\label{llm-context-recovery}

As discussed in Sec.~\ref{rollout_ft_background}, recovery latency includes engine restart, failure detection, and regeneration of interrupted requests.
To reduce this overhead, \SN supports LLM context recovery.

\SN stores the LLM context after each completed response. 
However, response-level recovery is insufficient for long reasoning segments with few environment interactions, such as math reasoning or code generation, where a single response may contain thousands of tokens. 
To provide finer granularity, \SN implements token-level recovery in the rollout router, following RLboost~\cite{wu2025rlboost}.
During streaming generation, the router incrementally records generated tokens and associated metadata, and updates the preserved prefix every \emph{k} tokens. 
In our implementation, we empirically set $k$ to 256.
Because handover usually completes within a few seconds, an interrupted request resumes on the new worker once it is ready rather than being sent to another healthy worker.
This avoids load imbalance and additional GPU-memory pressure on healthy workers.

\subsection{Hardware Failure Handling}

\SN treats a repeated failure on the shadow worker within a short post-handover window as evidence of a hardware fault.
It then marks the corresponding GPU unhealthy and excludes it from the rollout cluster.
\SN continues training on the remaining healthy GPUs and triggers an alert for manual intervention.
The rollout router distributes the affected requests evenly across other healthy engines and resumes them through token-level context recovery.

%% file: 5_Env_FT.tex
\vspace{-5pt}
\section{Environment Fault Tolerance}
\label{environment-fault-tolerance}

This section presents \SN's environment fault-tolerance design. 
We first define the failure model, recovery scope, and recovery goal.

\noindent \textbf{Failure model.}
We consider fail-stop failures during environment execution, such as container crashes, provided that the latest ready checkpoint remains accessible.
% \SN targets failures that make the current container incorrect to continue, but does not attempt to repair arbitrary application-level corruption inside a still-running process.

\noindent \textbf{Recovery scope.}
\SN provides prefix-consistent recovery for container-local environment state under fail-stop failures.
This state includes the container's file-system and runtime state as discussed in Sec~\ref{agentic_rl_background}.
\SN takes checkpoints at action boundaries, after the foreground environment action has returned.
% Its checkpoint primitive freezes all container processes while capturing the file-system and runtime components, so both originate from the same quiescent point.

\noindent \textbf{Recovery goal.}
Environment recovery returns a failed trajectory to a consistent prefix.
The recovered file-system state, runtime state, and LLM context should correspond to the same legal trajectory prefix, and subsequent observations should depend only on that prefix and actions regenerated after recovery.

\SN achieves consistent and low-interference environment recovery through an adaptive checkpointing mechanism (Sec.~\ref{environment-state-recovery}). 
% We also briefly describe an optional admission-control optimization used in our implementation (Sec.~\ref{sec:admission_optimization}).
\vspace{-5pt}
\subsection{Environment State Recovery}
\label{environment-state-recovery}

We follow two principles for environment-state recovery:
\begin{itemize}[itemsep=1pt, leftmargin=*]
\item \emph{Prefix consistency.} Recovery should restore file-system state, runtime state, and LLM context from the same ready trajectory prefix.
\item \emph{Low interference.} Recovery should impose minimal overhead on failure-free training.
\end{itemize}

Environment failures may occur at any point during an environment step.
A failure may leave the environment in a \textbf{corrupted state} from which the agent task cannot safely continue. 
Naively replaying environment actions from an ambiguous state is unsafe because many actions are non-idempotent: re-executing them may duplicate persistent effects or interact with a partially updated runtime, while dropping them may lose effects that were already observed by the LLM.

A more robust solution is to roll back to a consistent prefix and \textbf{regenerate} subsequent actions rather than replay the original actions verbatim.
This distinction matters because an action's effect may depend on nondeterministic inputs such as wall-clock time.
A conservative solution restarts the entire trajectory from the base image. 
Although this avoids reusing corrupted state, it is expensive for trajectories that run for minutes and exhibit a long tail, and it can delay completion of the rollout batch.

To avoid regenerating an entire trajectory from scratch, \SN uses checkpoint-based environment recovery. 
In general, an agent's environment state may span three components: containerized file-system state, container runtime state, and state in external systems.
\SN prevents agent actions from modifying external systems through sandboxing; therefore, the recoverable environment state within our scope is the pair of file-system and runtime state.
\SN introduces \texttt{full\_checkpoint} to materialize this complete container-local state and \texttt{full\_restore} to reconstruct it.
A full checkpoint is associated with an action-boundary trajectory prefix and becomes recoverable only after both state components have been captured successfully.
During recovery, \SN applies \texttt{full\_restore} to the latest ready full checkpoint and reloads the LLM context associated with the same prefix.
Figure~\ref{fig:full_state_api_usage} shows the corresponding API usage.

\begin{figure}[t]
\begin{lstlisting}[
language=Python,
basicstyle=\ttfamily\footnotesize,
keywordstyle=\color{blue!70!black}\bfseries,
stringstyle=\color{green!45!black},
commentstyle=\color{gray!75!black},
emph={full_checkpoint,full_restore,publish,latest_ready_prefix,continue_from},
emphstyle=\color{blue!70!black}\bfseries,
numbers=left,
numberstyle=\scriptsize\color{black},
numbersep=6pt,
xleftmargin=1.4em,
columns=fullflexible,
keepspaces=true,
showstringspaces=false,
breaklines=false,
aboveskip=0pt,
belowskip=0pt]
checkpoint = full_checkpoint(
    container=env,
    checkpoint_id=ckpt_id,
    state_root=state_root,
    docker_root=docker_root)
if checkpoint.status == "ready":
    publish(checkpoint.checkpoint_id, llm_context)

# After an environment failure
ckpt_id, llm_context = latest_ready_prefix()
restored_env = full_restore(
    checkpoint_id_or_path=ckpt_id,
    state_root=state_root,
    docker_root=docker_root,
    container_id=target_env)
continue_from(restored_env, llm_context)
\end{lstlisting}
\caption{Full-state recovery API usage.}
\label{fig:full_state_api_usage}
\Description{Python-style pseudocode showing full checkpoint creation, publication of a ready prefix, and full restore after an environment failure.}
\end{figure}

% \textbf{Action lifecycle and commit points.}
% \SN treats each environment action as a state-machine transition over the trajectory prefix.
% The lifecycle is \emph{GeneratedAction}, \emph{ActionDispatched}, \emph{ActionCompleted}, \emph{ObservationLogged}, \emph{CheckpointStarted}, and \emph{CheckpointReady}.
% An action first enters \emph{GeneratedAction} when it is produced by the LLM, then \emph{ActionDispatched} when \SN sends it to the container executor.
% After the foreground command exits, the executor records the exit code, stdout, stderr, and resource metadata, moving the action to \emph{ActionCompleted}.
% Once this observation is durably appended to the trajectory log, the action reaches \emph{ObservationLogged}, and the LLM context can advance to the next turn.
% Checkpoint creation then proceeds through \emph{CheckpointStarted} and \emph{CheckpointReady}.
% \emph{CheckpointStarted} invokes \texttt{full\_checkpoint}; only after both the file-system and runtime snapshots are complete does the state become \emph{CheckpointReady} and advance the recoverable-prefix pointer.
% This separation avoids claiming a general exactly-once RPC protocol for arbitrary environment effects.
% An observation-logged action is part of the logical trajectory, but only actions included in the latest ready full checkpoint are guaranteed to survive a fail-stop environment recovery without regeneration.
% Thus, \SN resolves ambiguity by recovering to a known prefix rather than by deduplicating arbitrary external side effects.

\textbf{Correctness invariant.}
Let $E_k=(F_k,R_k)$ denote the container-local environment state after legal trajectory prefix $k$, with file-system component $F_k$ and runtime component $R_k$.
Let $L_k$ denote the corresponding LLM context.
A ready full checkpoint $C_k$ maintains three invariants.
First, \emph{state agreement}: $F_k$ and $R_k$ are captured while the container is frozen at the same action boundary.
Second, \emph{atomic publication}: the recoverable-prefix pointer advances to $k$ only after both components of $C_k$ are complete; a failure during checkpoint creation leaves the previous ready checkpoint in effect.
Third, \emph{recovery alignment}: \texttt{full\_restore} reconstructs $E_k$, while the control plane restores $L_k$, so no subsequent observation can depend on partial effects from actions later than $k$.
These invariants follow from mutually exclusive action execution and checkpoint creation, delayed publication of composite checkpoints, and recovery in a fresh container rather than reuse of the failed instance.

However, frequent checkpointing can fall on the critical path of trajectory generation and interfere with training performance. 
To reduce this overhead, \SN adopts an adaptive risk-aware checkpointing strategy that jointly considers regeneration risk and natural LLM inference bubbles.

\textbf{Key observation.}
LLM responses naturally create bubbles between environment turns,
which provide opportunities to persist environment state without delaying execution.

Based on this observation, \SN formulates checkpointing as an \textbf{online cost-benefit decision} after each environment step. 
Once a command finishes, \SN decides whether to materialize the current container state as a recoverable checkpoint. 
As shown in Algorithm~\ref{alg:bubble_aware_checkpoint}, the decision compares the expected regeneration cost avoided by checkpointing with the visible checkpoint cost that cannot be hidden by the current LLM-response bubble.

\begin{algorithm}[t]
\caption{Adaptive Risk Checkpointing}
\label{alg:bubble_aware_checkpoint}
\begin{algorithmic}[1]
\Input Completed-step stream, latest checkpoint $c$, per-step failure-probability estimator $\hat{p}$, checkpoint-duration estimate $c_t$, conditional bubble-survival estimator $q$
\Output The latest ready checkpoint $c^\star$
\State $C \gets 0$, $c^\star \gets c$
\For{each completed step $s_t$}
    \State $C \gets C + \widehat{\mathrm{regenerate}}(s_t)$
    \State $p_t \gets \hat{p}(s_t)$
    \State $B \gets p_t \cdot C$
    \State $O \gets \int_0^{c_t} [1-q(x_t;u)]\,du$
    \If{$B \ge O$}
        \If{system is not busy}
            \State launch \texttt{full\_checkpoint}
            \If{checkpoint becomes \texttt{ready}}
                \State $c^\star \gets$ new checkpoint
                \State $C \gets 0$
            \EndIf
        \EndIf
    \EndIf
\EndFor
\State \Return $c^\star$
\end{algorithmic}
\end{algorithm}

Let $p_t$ denote the fail-stop probability during an environment step, estimated from historical data.
For example, under the standard constant-hazard exponential lifetime model~\cite{trivedi2002probability}, $p_t=1-e^{-\lambda d_t}$, where $\lambda$ is the fail-stop rate per unit time and $d_t$ is the wall-clock duration of step $t$.
For step-level failure-injection experiments, $p_t$ can also be set directly to the fixed injected failure probability.
Let $C_t$ denote the expected regeneration cost if recovery returns to the latest ready checkpoint.
In our implementation, $C_t$ is measured in wall-clock seconds and estimated from completed LLM responses and environment actions.
The expected benefit is the regeneration cost weighted by the probability of losing the current unprotected work:
\begin{equation}
B_t = p_t \cdot C_t.
\end{equation}
A larger $C_t$ means more work must be regenerated if the current container becomes unusable, while a larger $p_t$ means the next environment step is more likely to fail.

To compute checkpoint overhead, we model the LLM response time as a random variable $T$.
At decision point $t$, the current response has lasted $x_t$ seconds, so its remaining bubble is $R_t=T-x_t$, conditioned on $T>x_t$.
If the predicted checkpoint duration is $c_t$, launching it exposes only the portion that outlasts this bubble:
\begin{equation}
V_t=(c_t-R_t)^+,
\end{equation}
where $(z)^+=\max(z,0)$.
For any candidate duration $u$, we compute the conditional survival probability
\begin{equation}
q(x;u)=\Pr(T>x+u \mid T>x).
\end{equation}
This is the probability that the current LLM bubble lasts at least another $u$ seconds, conditioned on having already lasted $x$ seconds.
The survival curve can be estimated from the first one or two steps and updated as training dynamics change~\cite{zhou2025rlwildcharacterizingrlvr}.
The expected visible checkpoint cost is therefore
\begin{equation}
O_t
= \mathbb{E}\!\left[V_t \mid T>x_t\right]
= \int_0^{c_t}\!\left[1-q(x_t;u)\right]\,du.
\end{equation}
This expression accounts for partial overlap: if the response ends during checkpoint creation, only the checkpoint's remaining duration is visible.
Here, $c_t$ is predicted from checkpoint durations observed under the current load.
If the duration predictor instead represents checkpoint time as a random variable $D_t$, the corresponding general form is
\begin{equation}
O_t=\mathbb{E}\!\left[(D_t-R_t)^+ \mid T>x_t\right].
\end{equation}
% Our measurements use 13 seconds on the HDD setup and 6 seconds on the SSD setup.
% Because high concurrency increases disk and CPU contention, we use 28 seconds on the SSD setup in the scaling-to-frequent-failures experiments (Sec.~\ref{trajectory_state_recovery}).
The policy launches a checkpoint when the expected benefit exceeds the expected visible overhead:
\begin{equation}
B_t \ge O_t.
\end{equation}
The policy uses the current exposure time, estimated regeneration cost, elapsed LLM-response time, and checkpoint cost under current load.
If checkpoint creation times out, \SN disables checkpointing for a cooldown period, typically 20 seconds, to avoid repeatedly issuing checkpoints under heavy load.

\SN implements \texttt{full\_checkpoint} using Docker's layered storage and CRIU.
When a container is created, Docker places a writable upper layer over immutable image layers; this upper layer records container-local file-system changes while sharing unchanged data with the image~\cite{docker_storage_drivers}.
At an action boundary, \texttt{full\_checkpoint} first pauses all processes in the container.
While the container remains frozen, it archives the writable layer and invokes CRIU~\cite{criu} through Docker's checkpoint API~\cite{docker_docs} to dump the process tree and its runtime state into the same checkpoint directory.
\SN marks the composite checkpoint ready only after both artifacts have been stored successfully; otherwise, it discards the incomplete checkpoint and retains the previous recoverable-prefix pointer.

To recover, \texttt{full\_restore} terminates the failed container and creates a fresh container from the original image.
It installs the checkpointed upper layer into the new container's overlay directory and then invokes the \texttt{docker start} API with the corresponding runtime checkpoint.
Installing the file-system state before restoring the processes ensures that the resumed runtime observes the file-system image captured at the same action boundary.
For garbage collection, \SN discards the composite checkpoint directories at the end of the rollout.

%% file: 6_Cluster_FT.tex
\vspace{-5pt}
\section{Implementation}
\label{sec_implementation}
We implement \SN on top of Slime v0.2.1~\cite{slime_github}, a flexible and efficient LLM agentic RL training system with approximately 6K lines of code. 
The rollout cluster and shadow workers are implemented on top of SGLang v0.5.9~\cite{SGLang}, while the environment subsystem is built on mini-swe-agent~\cite{mini_swe_agent_github,yang2024sweagent}. 
For token level context recovery, we implement it in router with \emph{streaming} request handling.

\textbf{Shadow worker integration.}
After implementing the weight and KV cache servers (Sec.~\ref{shadow-worker}), we modify the rollout engine to support shadow worker.
SGLang's new \path{weight_daemon} loader receives model weights through CUDA IPC handles.
During initialization, \path{model_runner} calls \path{get_kv_cache_from_server} to obtain KV-cache pointers and metadata. 
We then integrate this shadow worker initialization path into \emph{Slime}.

\textbf{Full checkpoint/restore integration.}
We implement \texttt{full\_checkpoint} and \texttt{full\_restore} as a standalone module.
And then we integrate it into the rollout function to support environment checkpoint/restore once the checkpoint decision passed or environment failure occurs.
To support overlapping checkpointing with LLM generation, we launch the checkpointing process asynchronously. When the LLM generation finishes, we wait for the checkpointing process to finish if needed before proceeding.

%% file: 7_Evaluation.tex
\vspace{-5pt}
\section{Evaluation}
\label{evaluation}
Our evaluation addresses the following questions:
\begin{itemize}[itemsep=1pt, leftmargin=*]
\item \noindent \textbf{Interference-free execution.} Can \SN achieve end-to-end training performance comparable to a baseline system without fault-tolerance mechanisms? (Sec.~\ref{e2e_evaluation})
\item \noindent \textbf{Failure recovery.} Can \SN recover from different types and frequencies of failures with minimal training-time increase? (Sec.~\ref{rollout_failure_recovery} and Sec.~\ref{trajectory_state_recovery})
\end{itemize}

\subsection{Experiment Setup}
\label{exp_setup}

We evaluate \SN using Qwen3 series models~\cite{Qwen3} with 4B--32B parameters on two representative LLM agentic RL workloads. 
First, in the \emph{Math} workload, the model solves math problems through step-by-step reasoning; we use the dapo-17k-math dataset~\cite{DAPO}. 
Second, in the \emph{Software Engineering} (SWE) workload, the model fixes bugs in code repositories through iterative code inspection and modification; we use the SWE-Gym dataset~\cite{swe-gym}.

\textbf{Cluster setup.}
We deploy \SN on an Nvidia H200 GPU cluster with 4 nodes and 32 GPUs in total, connected by $400$~Gbps $\times$ 8 RoCE. 
For the environment cluster, we use an Ubuntu 24.04 machine with 16~CPU cores and 32~GB memory. 

\textbf{Models.}
For the Math workload, we evaluate Qwen3-4B, Qwen3-8B, and Qwen3-32B, with a maximum response length of 16K tokens. 
For the SWE workload, we evaluate Qwen3-32B due to model capability requirements, with a single step maximum response length of 4K tokens and a maximum context length of 32K tokens.

\textbf{Training configurations.}
We use GRPO~\cite{GRPO} as the training algorithm, with a rollout batch size of 64 and 8 samples per prompt. 
For the SWE workload, we set the maximum number of agent rounds to 20. 
We allocate three nodes to the rollout cluster and one node to the training cluster in 32 GPUs setting, and one node to each cluster in 16 GPUs setting. 
For parallelism, the 4B model uses TP=2 for rollout and $(\mathrm{DP}, \mathrm{TP})=(2,4)$ for training. 
The 8B and 32B models use TP=4 for rollout and $(\mathrm{DP}, \mathrm{TP})=(1,8)$ for training.

\textbf{Metrics.}
We use end-to-end training time as the primary metric for evaluating \SN's training performance and failure-recovery performance.

\subsection{End-to-End Evaluation}
\label{e2e_evaluation}

\begin{figure}[t]
    \centering
    \includegraphics[width=1\columnwidth]{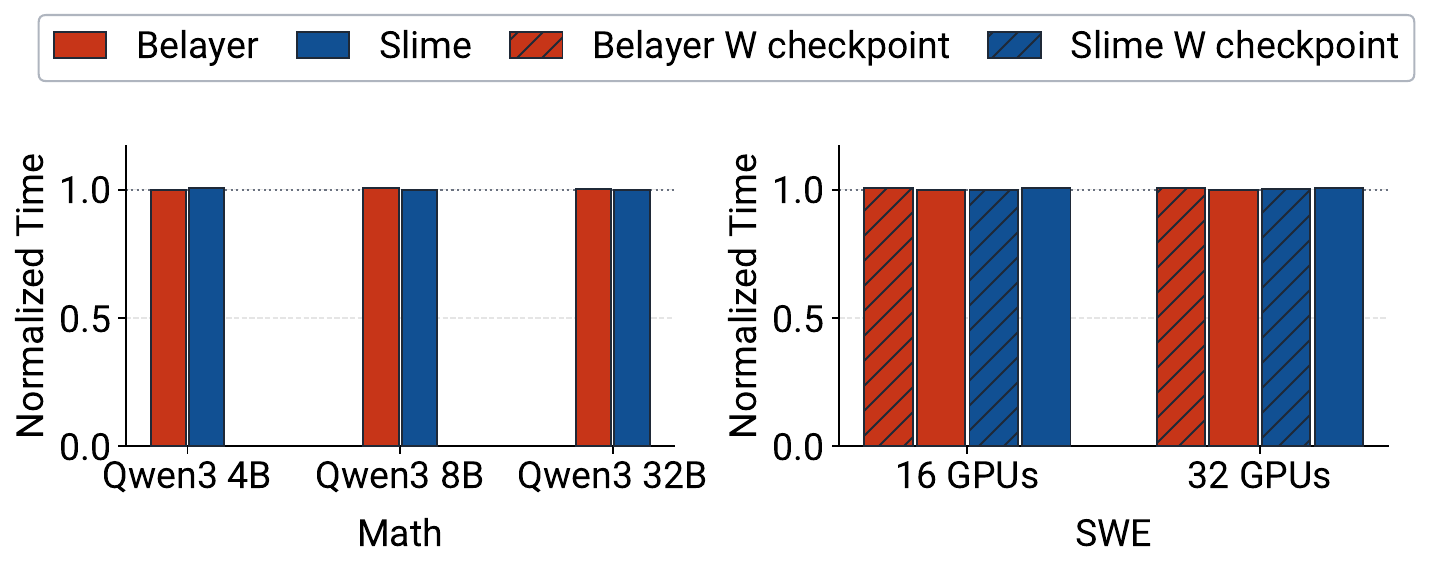}
    \caption{End-to-end training time of two workloads across three Qwen3 models.}
    \label{e2e_time}
\end{figure}

We first evaluate the end-to-end training performance of \SN and compare it with the baseline system to measure whether \SN's shadow workers and full-state environment checkpoints introduce overhead during failure-free execution. 
We do not compare against orthogonal optimization techniques, since such optimizations can be integrated with \SN to further improve performance.
For environment setting, we use the Kubernetes pod in this section.

% Because \emph{Slime} uses static environment concurrency, we tune its concurrency for a fair comparison. 
% We use the resource profile obtained in regular training process to find the reasonable static concurrency for \emph{Slime} that minimizes end-to-end training time. 
% The evaluated \SN configuration enables the lightweight admission-control optimization from Sec.~\ref{sec:admission_optimization}; we treat the result as a whole-system measurement and do not claim an isolated scheduling contribution.

As shown in Figure~\ref{e2e_time}, \SN achieves end-to-end training time comparable to the baseline. 
In the Math configurations, the measured training times differ by only 1\%. 
These measurements indicate low failure-free overhead in the evaluated configurations.
% For the SWE workload, \SN reduces training time by approximately 7\% compared with the baseline. 
% Because optional admission control is enabled in this configuration, we will discuss its effect in Sec.~\ref{workload_aware_scheduler}.
Moreover, there is little difference in training time between checkpoint and non-checkpoint scenarios.
Taken together, these results show that the combined system adds low failure-free overhead in the evaluated workloads.

\subsection{Rollout Worker Failure Recovery}
\label{rollout_failure_recovery}

We next evaluate \SN's instant restart mechanism under worker-local rollout failures during Math training.

\subsubsection{Recoverability}
We use real-world bug cases to evaluate \SN's recoverability.
We try to reproduce 17 SGlang issues(Sec.~\ref{opportunities}) in our testbed and successfully reproduce 8 of them due to lack of resources (some of which need large scale and specific hardware).
Table~\ref{tab:recoverability-examples} summarizes the reproduced cases.
We adapt our shadow worker mechanism in each issue's branch and verify that the shadow worker can take over the failed main worker and continue to serve requests.
For each case, we firstly obtain a deterministic baseline response in failure-free execution and
then reproduce the issue-specific failure and exits in main worker. The standby shadow detects the failure, discards the main worker’s
request-specific state and semantic KV contents, activates with fresh metadata, and resumes serving. 
All eight post-handover probes return HTTP 200, and deterministic recovered response match the pre-failure baseline request in generated text.

\begin{table}[t]
\centering
\normalsize
\setlength{\tabcolsep}{2.2pt}
\renewcommand{\arraystretch}{0.95}
\caption{Bug cases we reproduced and evaluated.}
\label{tab:recoverability-examples}
\begin{tabularx}{\columnwidth}{
  >{\raggedright\arraybackslash}p{0.25\columnwidth}
%   >{\centering\arraybackslash}p{0.2\columnwidth}
  >{\raggedright\arraybackslash}X
}
\toprule
\textbf{Issue No.}  & \textbf{Description} \\
\midrule
20640 & Queue sampling closure leak \\
19332 & Stale HiCache backing references \\
20010 & Mamba-pool slot-release omission \\
9888 & Stale CUDA-graph positions \\
11581 & KV allocator invariant failure \\
24912 & Uninitialized speculative bonus tokens \\
20485 & Cross-rank transfer-state skew \\
18980 & Speculative page-table extent mismatch \\

\bottomrule
\end{tabularx}
\vspace{-0.5em}
\end{table}

\subsubsection{Shadow Worker Overhead}
\begin{figure}[t]
    \centering
    \includegraphics[width=1\columnwidth]{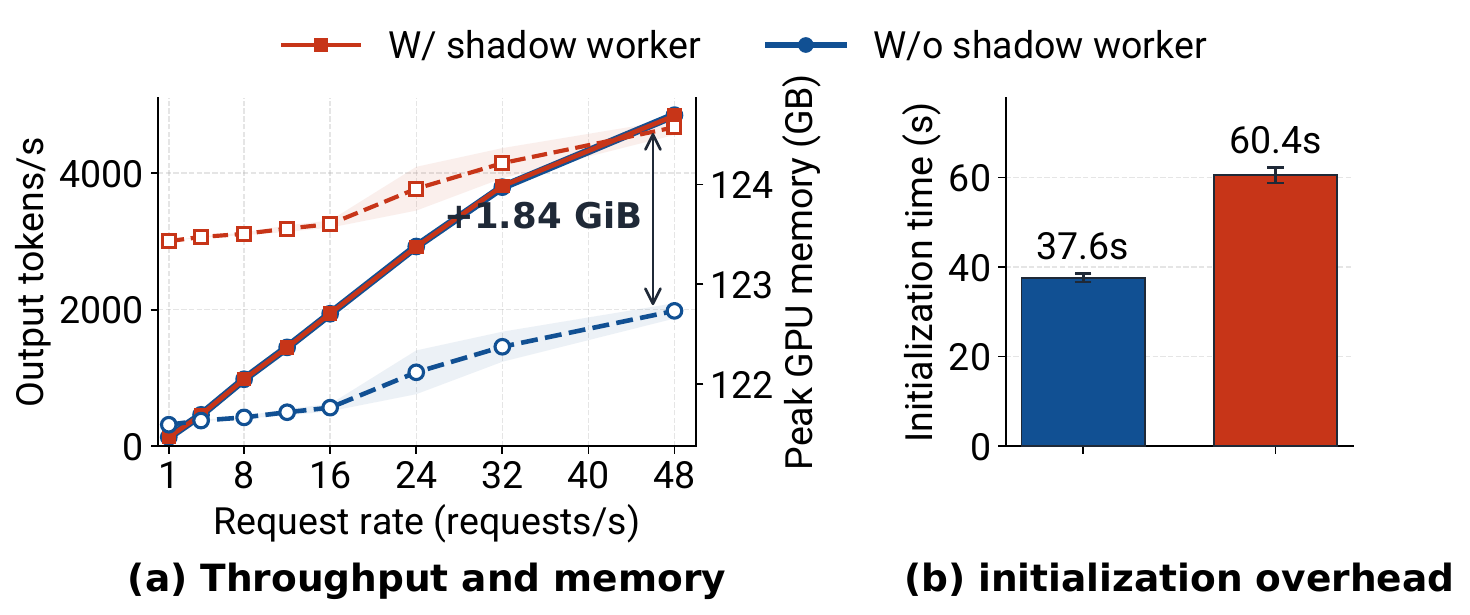}
    \caption{Colocation overhead of the shadow worker. We use Qwen3-32B with TP=4 on 4 GPUs to measure the overhead of the shadow worker. The dotted line indicates the memory usage while the solid one indicates the throughput.}
    \label{shadow_worker_overhead}
\end{figure}

As shown in Figure~\ref{shadow_worker_overhead}, the colocation of the shadow worker introduces minimal overhead.
For GPU memory, the shadow worker consumes only 1.84GB, including CUDA context and communication group memory consumption.
For the throughput, the shadow worker keeps the same throughput as the main worker. 
The shadow worker needs an extra initialization of process and CUDA context, which takes approximately 22.8 seconds. 
These results indicate that the shadow worker can be co-located with the main worker with acceptable resource overhead.

\subsubsection{Recovery efficiency}
We then measure how quickly \SN restarts a failed rollout worker. 

\noindent \textbf{Injection Method.}
We build a non-intrusive fault-injection tool by attaching runtime fault shims to the main worker while keeping the shadow worker unmodified. 
Python-level failures are injected through a \texttt{sitecustomize.py} hook that monkey-patches functions on the real inference path and raises native Python exceptions. 
CUDA and NCCL failures are injected through runtime hooks with \texttt{LD\_PRELOAD} that either return real CUDA/NCCL error codes or trigger CUDA allocation failures in the model forward path.

\noindent \textbf{Baseline.}
We use the RobustRL~\cite{Role-based_FT} approach as the baseline, which restarts the failed rollout worker from scratch. 
Instead of loading weights from storage, the new worker loads fresh model weights from healthy workers through high-speed interconnects. 
With this optimization, loading a Qwen3-32B model takes only about 5 seconds.

\begin{figure}[t]
    \centering
    \includegraphics[width=1\columnwidth]{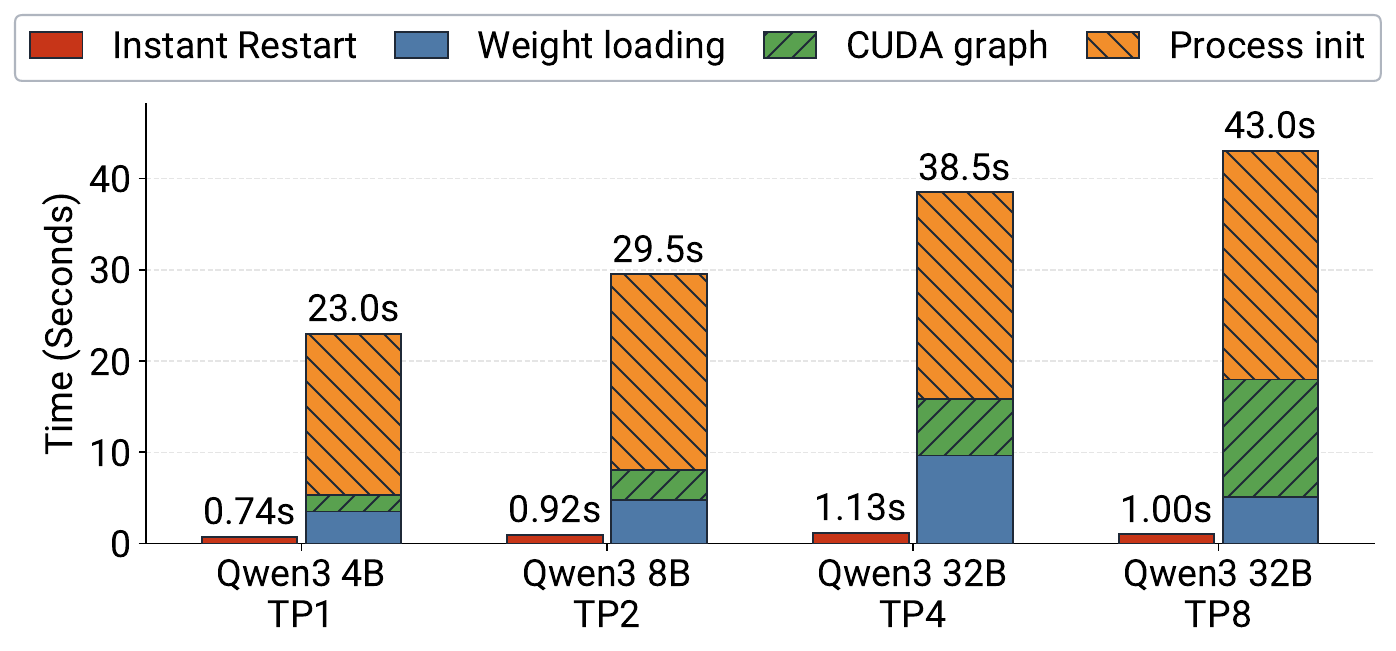}
    \caption{Restart time of rollout engines under different model sizes and TP sizes.}
    \label{instant_restart}
\end{figure}

As shown in Figure~\ref{instant_restart}, \SN restarts a failed rollout worker in approximately 1 second, whereas cold restart takes tens of seconds. 
This demonstrates the effectiveness of \SN's instant restart mechanism.
The cold-restart latency is dominated by process initialization. 
LLM inference engines must initialize many components, including the tokenizer, scheduler, HTTP server, torch context and so on, which together form a relatively fixed and expensive cost. 
Weight loading is comparatively faster because model weights remain resident on local disk to use faster PCIe transfers instead of loading from remote storage.
% Weight loading is comparatively smaller in our implementation because model weights remain resident on local disk after the first remote load, allowing subsequent restarts to use faster PCIe transfers. 
% For comparison, loading Qwen3-32B weights from remote storage, GPFS in our setup, takes 102.3 seconds.

\begin{figure}[t]
    \centering
    \includegraphics[width=1\columnwidth]{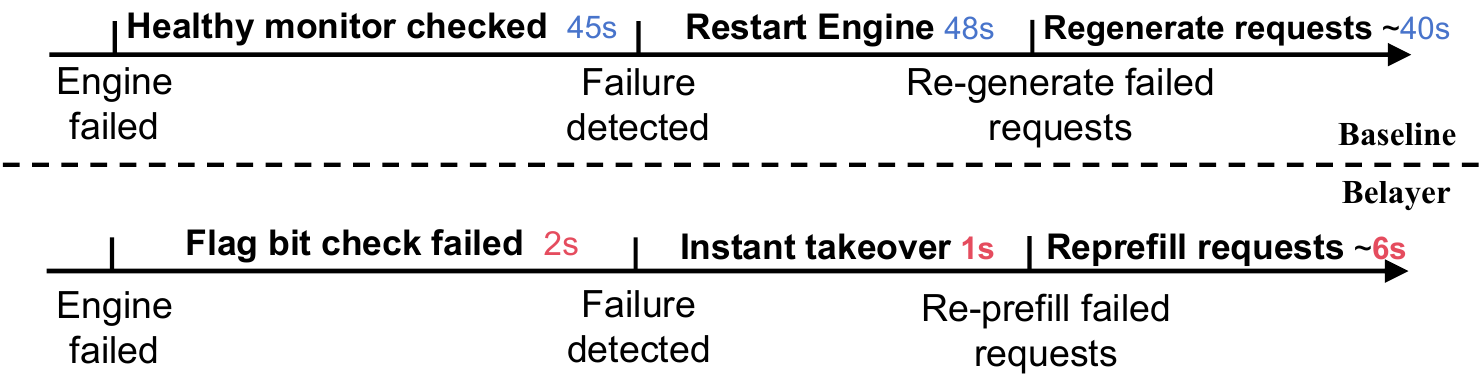}
    \caption{Breakdown of rollout-engine failure recovery time.}
    \label{fault_recovery_overview}
\end{figure}

\begin{figure}[t]
    \centering
    \includegraphics[width=1\columnwidth]{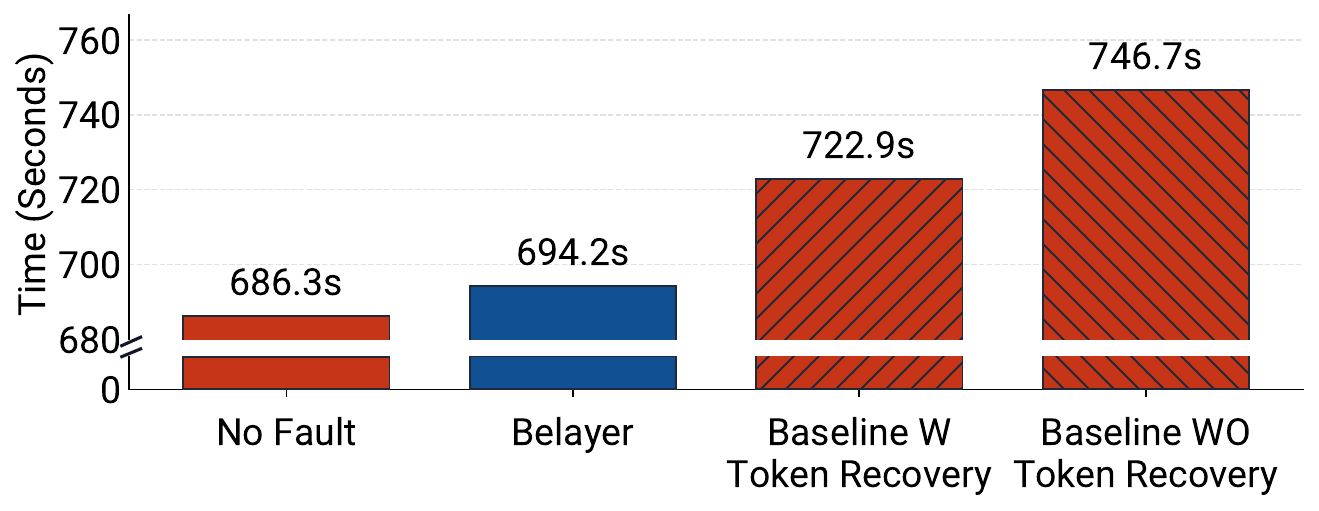}
    \caption{Single-step training time when a rollout engine encounters a software failure.}
    \label{fault_recovery_e2e_bar}
\end{figure}

We then evaluate the impact of rollout-worker failures on overall training time. 
As shown in Figure~\ref{fault_recovery_e2e_bar}, we inject a random software failure to a rollout worker process at the same point in the training process for both \SN and the baseline, and measure the resulting training time. 
\SN recovers from the software failure with only a 1.16\% increase in training time, whereas the baseline incurs an 8.72\% increase. 
Without instant restart and handover, the baseline must wait for failure detection and restart the failed worker from scratch; together, these steps take about 90 seconds and substantially increase training time.
Token-level recovery further reduces the training-time increase by 3.46\%. 
This highlights the importance of fine-grained context recovery for reducing the impact of rollout-worker failures.
The handover process breakdown is shown in Figure~\ref{fault_recovery_overview}. \SN can achieve fast detection, restart and request recovery, all sub-processes are highly optimized to minimize training-time increase.

\textbf{Scaling to frequent failures.}
To evaluate \SN under more frequent rollout-worker failures, we inject failures at the same point with different frequencies and measure training-step time. 
We use the Qwen3-8B model with TP=2 on 16 GPUs, corresponding to four rollout engines. 
As shown in Table~\ref{tab:failed_engine_recovery_transposed}, \SN maintains a small training-time increase as the number of failures grows, while the baseline degrades substantially.
The baseline overhead increases because it must restart more rollout workers from scratch. 
When more than two engines fail, the baseline exhibits a sudden increase in training time. 
This occurs because the number of failed engines exceeds the number of healthy engines available for remote weight loading, forcing some engines to fall back to loading weights from disk. 
In contrast, \SN's instant restart mechanism effectively mitigates the training-time impact even under frequent failures.

\begin{table}[t]
\centering
\caption{Rollout time under different numbers of failed engines. Numbers in parentheses indicate the increase over failure-free rollout time.}
\label{tab:failed_engine_recovery_transposed}
\setlength{\tabcolsep}{4pt}
\small
\resizebox{\linewidth}{!}{%
\begin{tabular}{c|c|c|c|c}
\hline
{\# of Failures} & {1} & {2} & {3} & {4} \\\hline
{Ours} & \makecell[c]{597.2s \\ (\textbf{+3.0s})} & \makecell[c]{600.3s \\ (\textbf{+6.1s})} & \makecell[c]{611.5s \\ (\textbf{+17.3s})} & \makecell[c]{620.4s \\ (\textbf{+26.2s})} \\\hline
{\makecell{Baseline \\ W Token Recovery}} & \makecell[c]{614.7s \\ (\textbf{+20.5s})} & \makecell[c]{618.6s \\ (\textbf{+24.4s})} & \makecell[c]{691.7s \\ (\textbf{+97.5s})} & \makecell[c]{695.6s \\ (\textbf{+101.4s})} \\\hline
\end{tabular}%
}
\end{table}

\subsection{Environment State Recovery}
\label{trajectory_state_recovery}

\begin{figure}[t]
    \centering
    \includegraphics[width=1\columnwidth]{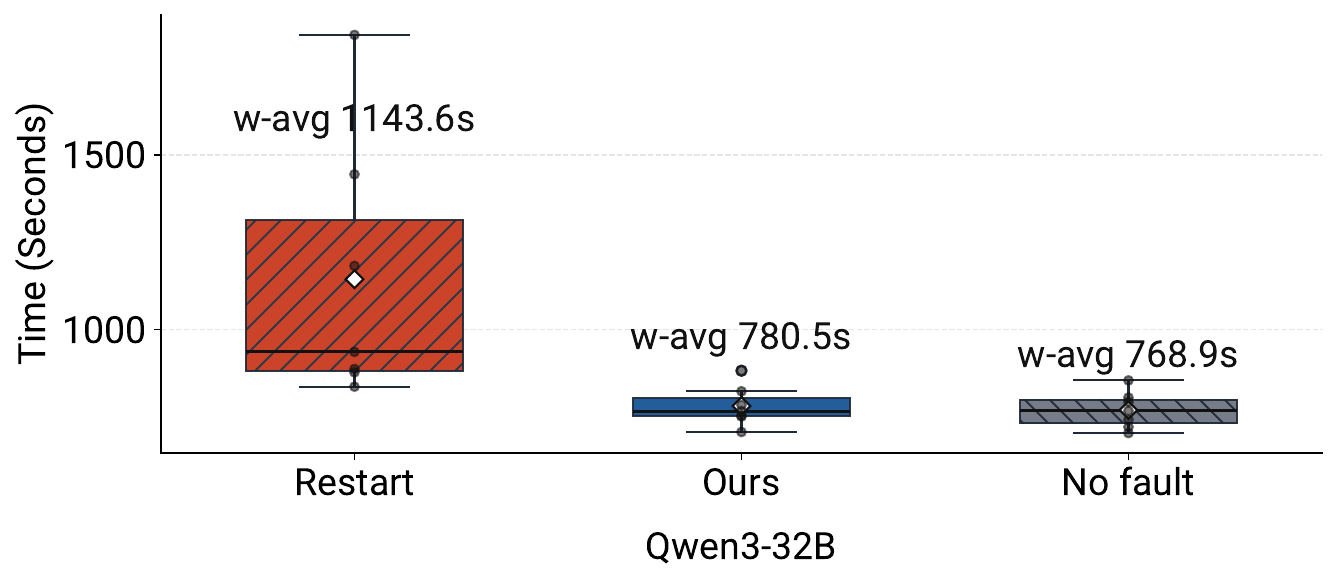}
    \caption{Rollout time under a faulty environment cluster. We measure the rollout time of 128 trajectories over multiple runs using a single rollout engine with Qwen3-32B and TP=4.}
    \label{trajectory_fault_recovery_boxplot}
\end{figure}

In this subsection, we evaluate \SN's prefix-consistent environment-state recovery based on \texttt{full\_checkpoint} and \texttt{full\_restore}.
We use Pumba~\cite{ledenev-pumba}, a chaos-testing tool for Docker, to inject fail-stop failures into environment containers with probability 0.01 per execution step and measure total rollout time. 
Because both failures and LLM responses are stochastic, we repeat the rollout multiple times. 
As shown in Figure~\ref{trajectory_fault_recovery_boxplot}, \SN substantially reduces the rollout-time increase caused by environment-container failures compared with the baseline.

\subsubsection{Recovery correctness}

\textbf{Methodology.}
We apply differential testing with a golden-run oracle~\cite{McKeeman1998Differential}.
We use three independently synthetic action sequences, for each sequence, a fault-free container executes six actions to produce reference runtime and file-system states.
After a fail-stop, the system restores the latest \texttt{ready} checkpoint and replays the remaining actions; if none exists, it restarts from the initial state.
A trial passes only when both final states match the corresponding reference.

The six actions jointly exercise persistent and live state: PRNG and accumulator states, an open-file offset, an ordered token transcript, selected bytes in a 16~MiB heap, and file creation, rename, deletion, symbolic-link creation, permission changes, directory creation, and file moves.
Cross-coupled runtime and file-system hash chains expose inconsistent combinations.

For arbitrary time injection, we divide each seed's fault-free duration into 200 equal-width strata, preselect one random wall-clock offset per stratum, and inject a fail-stop independently of execution phase (600 trials total).
% Phase traces are used only for post-hoc coverage classification and cannot influence injection.
% A separate systematic campaign targets all 42 hooks (9 action, 16 checkpoint, and 17 restore), six internal windows, and two base-restart cases.

For runtime equivalence, both executions serialize logical progress, PRNG and accumulator states, the actual open-file offset, the action transcript, and all 16~MiB heap bytes into the same canonical format.
After validating their structure, we compare the artifacts byte for byte using GNU \texttt{cmp}~\cite{GNUcmp}.
Following reference-state comparison in crash-consistency testing~\cite{Mohan2018CrashConsistency}, we check file-system equivalence with a checksum-based \texttt{rsync} dry run~\cite{rsyncManual} over paths, contents, symbolic and hard links, permissions, ownership, ACLs, and extended attributes.
A zero exit status with no itemized changes indicates a match.

\textbf{Results.}
Of the 600 scheduled arbitrary-time trials, 593 interrupted active executions, and all recovered to the golden final state: 530 restored a ready checkpoint, while 63 restarted from the initial state because none was ready.
The remaining seven signals arrived after completion.
% The active faults sampled 20 of 32 execution intervals; the 12 unsampled intervals were short.
% Moreover, the systematic campaign covered all 42 hooks (42/42), six internal windows, and both base-restart cases.
% Thus, 20/32 measures temporal coverage under arbitrary-time injection, whereas 42/42 measures boundary coverage under directed injection.

\textbf{Conclusion.}
Across 593 active arbitrary-time fail-stops trials, we observed no consistency counterexample.
Within the tested scope, the protocol discarded incomplete checkpoints, restored the latest ready state (or the initial state), replayed the remaining actions exactly once, and converged to the fault-free runtime and file-system state.

\subsubsection{Recovery efficiency}
To the best of our knowledge, no existing system provides a detailed design and implementation for environment-state recovery under environment failures in LLM agentic RL training. 
We therefore compare \SN against restart from scratch and two fixed policies that invoke \texttt{full\_checkpoint} after every action or every three actions.

\begin{figure}[t]
    \centering
    \includegraphics[width=1\columnwidth]{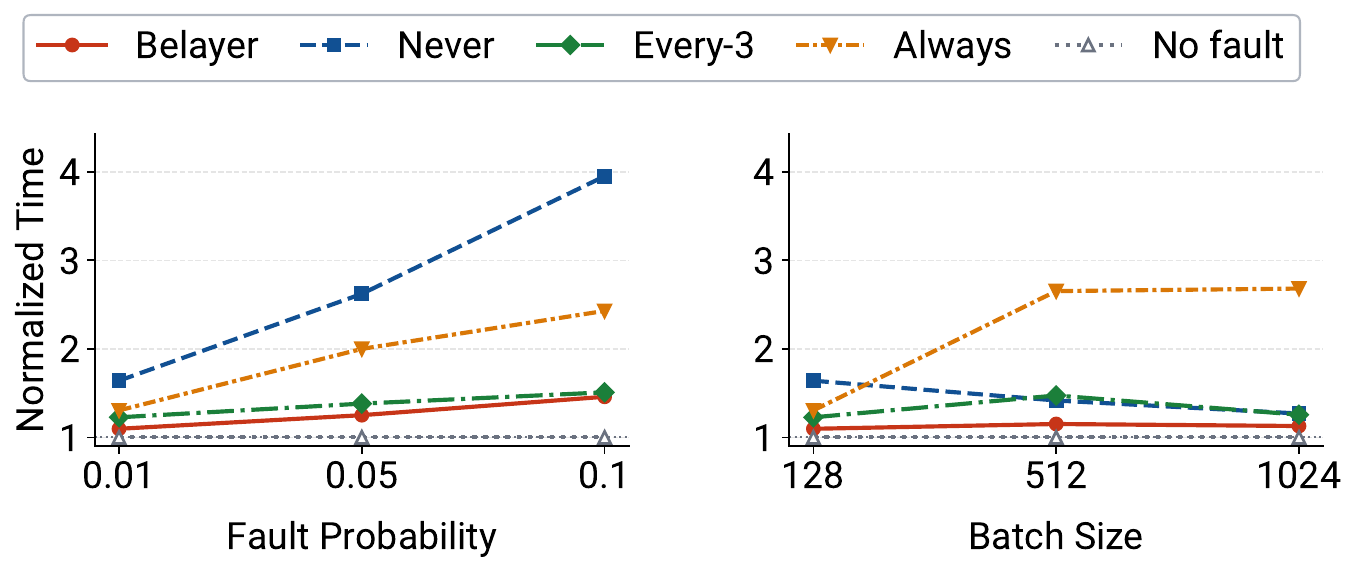}
    \caption{REDO: Normalized rollout time under a faulty environment cluster with different failure rates and batch sizes.}
    \label{trajectory_fault_recovery_normalized_lines}
\end{figure}

Restarting from scratch without environment-state recovery increases rollout time by 48.7\% on average because the failed trajectory can block rollout progress. 
With full-state recovery, \SN incurs 1.5\% rollout-time increase. 
Restoring the latest ready checkpoint selected by \SN's adaptive policy avoids regenerating the trajectory from the base image and substantially reduces recovery latency.

\textbf{Scaling to frequent failures.}
We further evaluate the scalability of \SN's environment-state recovery through simulation. 
We deterministically replay a group of trajectories to simulate the rollout process. 
The LLM response is abstracted as a \emph{sleep} operation recorded in each trajectory, while the environment component uses the same real environment and replays the recorded commands at each step. 
As shown in Figure~\ref{trajectory_fault_recovery_normalized_lines}, the baseline rollout time increases significantly as the failure rate grows, whereas \SN maintains a small rollout-time increase. 
This is because \SN restores failed environments from recent full checkpoints and avoids expensive regeneration from the base image. 

\textbf{Scaling to larger batches.}
We also evaluate environment-state recovery under larger rollout batches with a fixed failure rate of 0.01. 
As shown in Figure~\ref{trajectory_fault_recovery_normalized_lines}, environment failures have less impact on rollout time as batch size increases. 
With larger batches, the rollout process has more trajectories to execute, allowing the cost of a single failed trajectory to be overlapped with other ongoing trajectories. 
In contrast, frequent full-state checkpointing can significantly increase checkpoint overhead and degrade performance.
\SN continues to recover failed trajectories effectively, with only slight performance degradation.

\begin{table}[t]
\centering
\normalsize
\caption{REDO: Full-checkpoint overhead under different storage backends when generating $16 \times 8$ trajectories with a failure rate of 0.01 per step.}
\label{tab:checkpoint-overhead-backend}
\setlength{\tabcolsep}{3.5pt}
\begin{tabular*}{\columnwidth}{@{\extracolsep{\fill}}llrrrrr}
\toprule
\textbf{Storage} & \textbf{Policy} & \textbf{E2E(\%)} & \textbf{\#Ckpt} & \textbf{Total(s)} & \textbf{Exposed(s)} \\
\midrule
HDD & Always   & 276.7 & 1143 & 45122.3  & 3130.6 \\
HDD & Every-3  & 44.4  & 243  & 3989.1   & 188.6 \\
HDD & Ours & \textbf{21.6}  & \textbf{46}   & \textbf{641.7} & \textbf{20.3} \\
\midrule
SSD & Always   & 12.5 & 674 & 4974.7 & 98.0 \\
SSD & Every-3  & 6.3  & 209 & 1137.3 & \textbf{9.7} \\
SSD & Ours & \textbf{2.1}  & \textbf{179} & \textbf{967.0} & 14.0 \\
\bottomrule
\end{tabular*}
\vspace{0.3em}
\begin{minipage}{0.98\columnwidth}
\footnotesize
The E2E column shows the end-to-end training time increase compared with the oracle-no-fault-no-checkpoint policy. 
The Total column shows the total time spent on full-checkpoint creation, 
and the Exposed column shows the portion of that time not hidden behind LLM response latency.
\end{minipage}
\end{table}

\textbf{Full-checkpoint overhead analysis.}
Table~\ref{tab:checkpoint-overhead-backend} shows that full-checkpoint overhead is highly sensitive to the storage backend. 
On the HDD VFS setup, invoking \texttt{full\_checkpoint} after every action is prohibitively expensive: 
it creates 1,143 checkpoints, spends 45.1K seconds in checkpoint creation across trajectories, and increases E2E runtime by 276.7\%. 
Checkpointing every three steps reduces the exposed batch overhead to 188.6 seconds, but still increases E2E runtime by 44.4\%. 
In contrast, adaptive-risk creates only 46 checkpoints, hides 96.8\% of checkpoint time behind LLM-response latency, and leaves only 20.3 seconds exposed on the critical path, reducing E2E overhead to 21.6\%.
With the SSD overlay2 setup, full-checkpoint creation is substantially cheaper, so \SN creates more checkpoints to reduce regeneration overhead,
increasing total runtime by only 2.1\%, compared with 6.3\% for Every-3 and 12.5\% for Always. 
These results show that \SN adapts the frequency of full-state checkpoints to storage cost and recovery risk.

%% file: 8_discussion.tex
\section{Discussion}
\label{discussion}

\textbf{Limitations.} \SN has three limitations.
First, its environment recovery primitives rely on Linux containers, Docker's overlay storage, and CRIU; supporting virtual machines or other isolation backends would require analogous storage and runtime snapshot mechanisms.
Second, \SN provides prefix-consistent recovery for container-local file-system state and CRIU-restorable runtime state, but not exactly-once semantics for effects outside the sandbox. Its guarantee excludes external-service state, bind-mounted volumes, host-side daemons, device state, and live connections whose peers are not included in the checkpoint.
Third, \SN's shadow-worker handover targets worker-local software failures and does not guarantee warm recovery from CUDA context corruption, driver resets, GPU hardware faults, or failures of the weight/KV-cache servers.

\textbf{Future work.} We plan to investigate more optimistic recovery strategies and extend full-state recovery to environments that interact with external services.
Other directions include flush-free KV-cache recovery for shadow workers and hardware-failure handling through techniques such as live migration and elastic rollout scaling.

\section{Related Work}
\label{related_work}

\textbf{LLM RL training systems.}
Early efforts on LLM RL training~\cite{verl,Puzzle,Openrlhf,deepseed_chat} aim to orchestrate the complex training workflows of LLM RL training.
Many works further optimize the training efficiency under long-tail rollout workloads.
These works~\cite{Areal,laminar,StreamRL} optimize system efficiency by asynchronous RL training.
RLHFuse\cite{RLHFuse} focuses on optimizing RLHF training through stage fusion,
Rollpacker~\cite{Rollpacker} optimizes rollout efficiency through long tail batching,
Seer~\cite{Seer} uses online context learning for fast LLM RL training.
Moreover, speculative decoding mechanisms~\cite{TamingTheLongTail,RhymeRL,RLspec} are used to tame the long-tail workloads.
However, none of these works provide efficient and correct failure recovery mechanisms for the distributed execution model of LLM agentic RL training.

\textbf{Fast failover and LLM cold start.}
Primary-backup handover and pre-initialized standby workers reduce failover latency by keeping replacement capacity ready~\cite{Primary_backup,Chubby}.
Partial-state recovery instead shortens restart by preserving selected state across a failure~\cite{OptimisticRecovery}.
LLM-serving systems specialize cold-start optimization for heavyweight inference engines.
ServerlessLLM accelerates weight loading and coordinates startup-aware model placement~\cite{ServerlessLLM}; Medusa materializes initialized runtime state, including CUDA graphs~\cite{MedusaMaterialization}; and HydraServe proactively distributes models, overlaps cold-start stages, and places workers to reduce contention~\cite{HydraServe}.
These systems reduce different portions of serverless instance startup, but replacing an inference-engine instance still incurs the portions of initialization that are not retained or overlapped.
\SN combines pre-initialized shadow workers with independently owned GPU allocations, post-failure owner and GPU health checks, request-specific KV invalidation, and token-prefix reconstruction to recover an already-running agentic RL rollout engine.

\textbf{LLM training fault-tolerance.}
Checkpointing model states are widely used for fault-tolerance in DNN training~\cite{strati2025pccheck,lian2025universal,ByteCheckpoint,CheckFreq}.
Moreover, other efforts~\cite{Varuna,Parcae,Bamboo,Oobleck,ma2026resihp,Recycle} can training LLM in elastic clusters with dynamic resource availability.
However, these works mainly focus on training phase, and do not consider the unique challenges of LLM agentic RL training, such as trajectory consistency.

%% file: 9_Conclusion.tex
\section{Conclusion}

This paper presents \SN, a fault-tolerant and efficient system for LLM agentic RL training that
provides fast rollout recovery and prefix-consistent recovery of container-local environment state through composite file-system and runtime checkpoints. 
Our experiments show that \SN can significantly reduce recovery time with low failure-free overhead under the evaluated workloads.

%% file: Appendix.tex
\begin{comment}
\appendix

\section{Supplementary Admission-Control Results}
\label{appendix:admission_results}

Figure~\ref{exec_server_resource_timeseries} provides the resource-utilization and running-container traces behind the aggregate results in Table~\ref{tab:scheduler_static_baselines}.
All policies replay the same $16 \times 8$ trajectories; the figure is included as supporting characterization of the optional implementation optimization.

\begin{figure}[H]
    \centering
    \includegraphics[width=1\columnwidth]{scheduler_static_resource_appendix.pdf}
    \caption{Environment resource utilization and running container count over time.}
    \label{exec_server_resource_timeseries}
\end{figure}
\end{comment}